\documentclass[10pt,aps,pre,twocolumn,superscriptaddress]{revtex4-2}
\usepackage[linkcolor = blue, citecolor = red, urlcolor = blue, colorlinks = true]{hyperref}
\usepackage[usenames,dvipsnames,x11names]{xcolor}
\usepackage[normalem]{ulem}
\usepackage{graphicx}
\usepackage{listings}
\usepackage{stmaryrd}
\usepackage{amssymb}
\usepackage{amsmath}
\usepackage{gensymb}
\usepackage{comment}
\usepackage{float}
\usepackage{bbold}
\usepackage{bm}
\definecolor{crimson}{rgb}{0.75686,0,0.262745}
\definecolor{saphire}{rgb}{0.0,0.196,0.372549}
\definecolor{plum}{rgb}{0.50588,0.007843,0.3843137}

\usepackage[english]{babel}
\usepackage{lmodern}
\usepackage[utf8]{inputenc}
\usepackage{endnotes}
\usepackage{amssymb,amsmath,amsfonts}
\usepackage{textcomp}
\usepackage{braket}
\usepackage{ulem}
\usepackage{soul}
\usepackage{graphicx}
\usepackage{dcolumn}
\usepackage{bm}
\usepackage{xcolor}
\usepackage{soul}
\usepackage{xr}
\usepackage{cleveref}

\newcommand{\fig}[1]{Fig.~\ref{#1}}

\newcommand{\tocite}[1]{\textcolor{blue}{[c]}}

\begin{document}

\title{Topological delocalisation of confined 3D active nematics}
\author{Louise C. Head}
\affiliation{Department of Physics and Astronomy, Johns Hopkins University, Baltimore, MD
21218, USA.}
\author{Pasquale Digregorio}
\affiliation{Dipartimento Interateneo di Fisica, INFN, Universit\`{a} degli Studi di Bari Aldo Moro, Bari, Apulia, Italy}
\author{Davide Marenduzzo}
\affiliation{SUPA, School of Physics and Astronomy, University of Edinburgh, Peter Guthrie Tait Road, Edinburgh, EH9 3FD, UK}
\author{Ignacio Pagonabarraga}
\affiliation{Departament de F\'{i}sica de la Materia Condensada, Universitat de Barcelona, Mart\'{i} i Franqu\'{e}s 1, E08028 Barcelona, Spain}
\affiliation{UBICS University of Barcelona Institute of Complex Systems, Mart\'{i} i Franqu\'{e}s 1, E08028 Barcelona, Spain}
\author{Daniel A. Beller}
\email{corresponding author: d.a.beller@jhu.edu}
\affiliation{Department of Physics and Astronomy, Johns Hopkins University, Baltimore, MD
21218, USA.}
\author{Giuseppe Negro}
\email{corresponding author: giuseppe.negro@ed.ac.uk}
\affiliation{SUPA, School of Physics and Astronomy, University of Edinburgh, Peter Guthrie Tait Road, Edinburgh, EH9 3FD, UK}


\begin{abstract}
    Defect lines in 3D active nematic systems are intriguing topological singularities whose out-of-equilibrium dynamics remain elusive in confined settings. Here, we numerically study 3D active nematics confined within closed cylinders to elucidate the roles of geometry and activity. 
    We reveal a competition between passive elasticity, which causes localisation of defects near edges, and activity, which endows defects with motility and gives rise to disorderly, delocalised dynamics. 
    Varying boundary curvature, activity strength, and cylinder radius reveals a state space of static and dynamic localisation states, including handle-like configurations and chaotic motion bounded within the cylinder endcap. As activity is tuned to induce delocalisation, we identify phase transition signatures, including pronounced fluctuations and an emergent power law scaling of defect number and average defect length. We find that these scaling properties are strongly altered by confinement: unlike in bulk systems where activity governs length distributions, confinement tunes an activity-independent characteristic length, with an exponent reminiscent of self-avoiding confined polymers. These results establish confinement of inhomogeneous curvature as a versatile mechanism for controlling active topological dynamics.

\end{abstract}

\maketitle

\section{Introduction}


Active nematics~\cite{marchetti,doostmohammadi2018} are fascinating non-equilibrium systems which encompass microtubule-motor mixtures~\cite{sanchez2012}, actomyosin suspensions~\cite{kumar2014,tjhung}, bacteria colonies~\cite{yashunksy2024}, cell monolayers~\cite{armengol2023,Armengol2024,chiang2023} and cancer tissues~\cite{argento_gliomas_2025}. 
In these systems, continual energy dissipation at the microscopic level gives rise to emergent, large-scale phenomena such as active turbulence~\cite{Alert2022}. 
This spontaneously flowing state has been extensively studied in 2D active films, where the persistent creation and annihilation of motile topological defects drive and sustain active turbulence~\cite{Giomi2015}. 
In 3D active nematics~\cite{duclos2020}, active turbulence is instead understood through the dynamics of disclination lines, whose spatially-extended configurations exhibit percolation and entanglement, reminiscent of polymer melts, with lines acting as dynamic motile filaments~\cite{Digregorio2024}. 



Introducing confinement, however, fundamentally alters this landscape. This is because, even in the passive limit, surface geometry and topological constraints dictate defect morphology. 
Topological constraints in active systems have largely focussed on 2D films on closed surfaces, where for spheres, the director is unable to align tangentially without introducing defects. Accordingly, intriguing defect behaviours arise with activity, including the periodic dynamics of $+1/2$ defects~\cite{keber2014,zhang2016} and protrusion-like phenomena in deformable surfaces, which has biological relevance to epithelial morphogenesis~\cite{Hoffmann2022} and hydra regeneration~\cite{maroudas2021}. Geometry, too, plays a significant role in tuning defect behaviour. Non-uniform surface curvature controls defect positioning, driving attraction towards ellipsoidal poles~\cite{Alaimo2017} and the apexes of cones~\cite{vafa2024}. However, little attention has been given to 3D active nematics confined within closed surfaces.
Instead of surface defects, defect lines connect between points on the boundary and drive bulk dynamics~\cite{copar2019,Nejad2023}. 
It remains to be investigated how geometry and activity can be harnessed to control disclination lines. Furthermore, while defect lines have been characterised in bulk~\cite{Digregorio2024,kralj_chirality_2024}, their properties under confinement are largely unknown.


\begin{figure*}[ht!]
  \centering
  \includegraphics[width=1.0\linewidth]{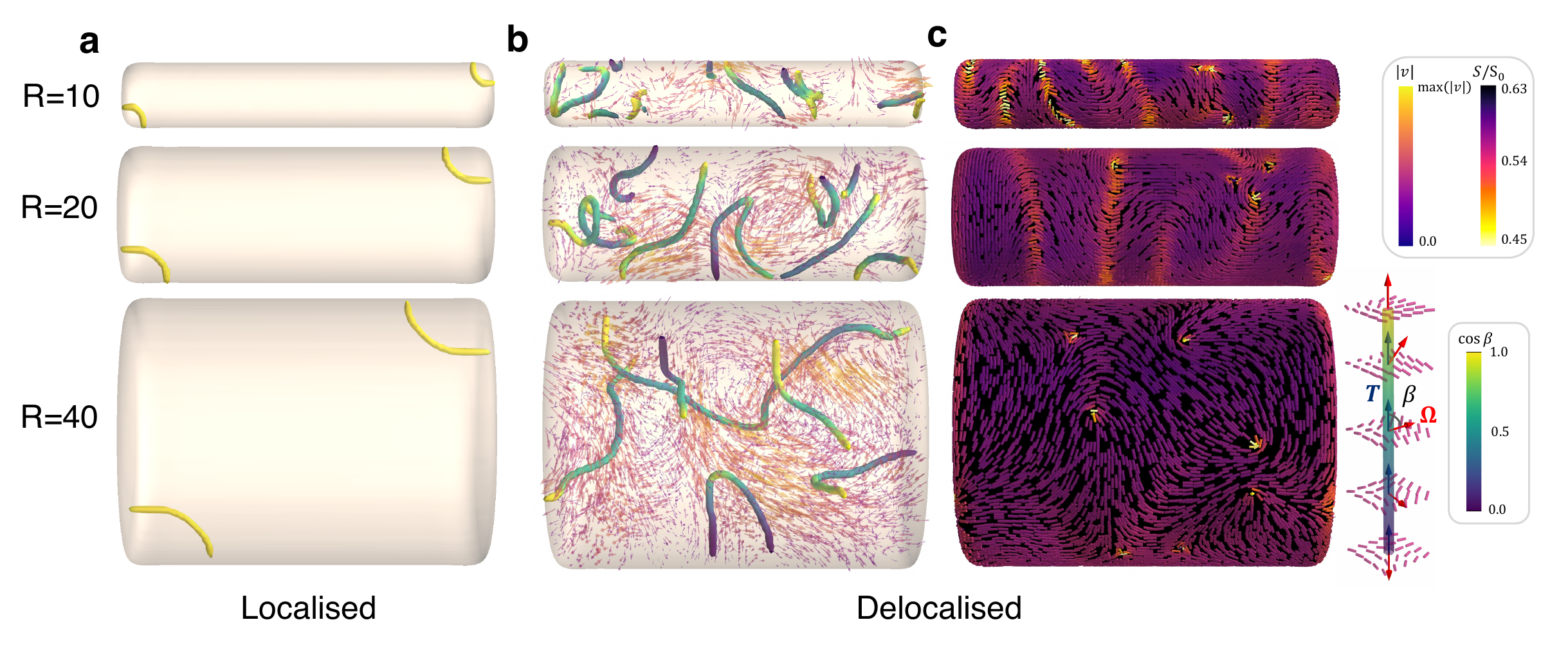}
\caption{\textbf{Snapshots of the localised and delocalised states.} \textbf{a}, Defects are localised near cylinder edges at equilibrium. Snapshots have activity number $\mathrm{A}=0$ and cylinder radii $R=10,20$ and $40$. \textbf{b}, Delocalised defects are mobile within the cylinder. Flow field shown as arrows and coloured by velocity magnitude $|v|$. 
Snapshots have $\mathrm{A}=15$ for $R=10$ and $\mathrm{A}=20$ for $R=20$ and $30$. Defect lines are coloured by $\cos\beta$, with local profile resembling $+1/2$ ($\cos\beta=1$; yellow), $-1/2$ ($\cos\beta=-1$; purple), or intermediate twist ($\cos\beta=0$; green/blue).
\textbf{c}, Same delocalised snapshot as \textbf{b} but showing director field on surface of cylinder. Director coloured by scalar order parameter $S$ scaled by the equilibrium value $S_0$.
}
\label{fig:localiseddelocalised}
\end{figure*}

In this work, we numerically investigate 3D active nematics enclosed within cylinders as a prototypical case of surfaces with non-uniform curvature. 
The director is planar anchored at the cylinder boundary, which is experimentally realisable since extensile in-plane flows naturally align the director with the surface~\cite{opathalage2019,blow2014,ruske2021}.
We investigate the transition from localised states, where defects are pinned to curvature maxima at the cylinder edges, to a regime of topological delocalisation associated with active turbulence. 
At activities near the onset of delocalisation, we uncover a surprisingly rich variety of defect behaviours. In addition to pinning at curvature extrema, defects can adopt handle configurations or undergo spatially confined fluctuations at the endcaps.  The emergence of such states is governed by the interplay between activity and boundary curvature. 
Intriguingly, the transition between localised and delocalised defect networks is qualitatively  reminiscent of the transition between topological insulators and topological superconductors~\cite{bernevig2013}, where flux lines (instead of disclinations) are either pinned to the surface or ``escape'' to permeate the bulk. However, we reveal that strong confinement can switch the delocalisation mechanism to defect nucleation along the channel. This mechanism is similar to systems without topologically required defects such as in bulk and in open channels~\cite{Chandragiri2020}, where defect nucleation is instability-driven~\cite{martinez2019}. 

Finally, we characterise the properties of defects within the delocalised state. By examining the internal geometry of local defect profiles, we distinguish between twist, comet and trefoil patterns~\cite{copar2013,copar2014,head2024-3}. Due to the competition between geometry and activity, these profiles localise to different regions of the cylinder, with comet profiles (with $+1/2$ topological charge) attracted within a localisation length from the endcaps.
We further show how defect lines proliferate in confined turbulence, and that their number scales with activity differently from defect loops, which dominate in bulk.
Lastly, we explore how delocalised defects behave as ``living polymers''~\cite{cates1987} within the cylinder. By analysing the defect contour lengths, we show that the probability of observing disclinations of size $l$ decays exponentially in confinement, at variance with the power law behaviour found in the bulk~\cite{Digregorio2024}. Additionally, we identify scaling regimes reminiscent of the de Gennes blob theory~\cite{deGennes1979}, whereby the average defect length scales with confinement radius as $l \sim R^{5/3}$, consistent with a self-avoiding random walk conformation.

\section{Results}

\subsection{Localised and delocalised states}

We begin by establishing the role of the cylindrical geometry in the spatial localisation of defects. The methods are discussed in the \textit{Methods} section; we briefly outline the set-up here. We numerically solve the continuum equations of active nematohydrodynamics using a hybrid lattice Boltzmann approach \cite{marenduzzo2007} with phase fields that specify the cylindrical active nematic domain and the outer isotropic region. 
The phase field representing the active nematic is initialised as a perfect cylinder and evolved with Cahn Hilliard dynamics for a short time to form rounded edges. Then, the liquid crystal is initialised with randomised orientations before being evolved in time to find the free energy minimum. 
Once the equilibrium configuration is found, the simulation evolves with the full active nematohydrodynamics with active stress $-\zeta \mathrm{Q}$, where $\mathrm{Q}$ is the nematic tensor order parameter and $\zeta$ is the phenomenological activity parameter. In this work we only consider extensile materials where $\zeta$ is positive. 
\begin{figure}
  \centering
  \includegraphics[width=1.0\linewidth]{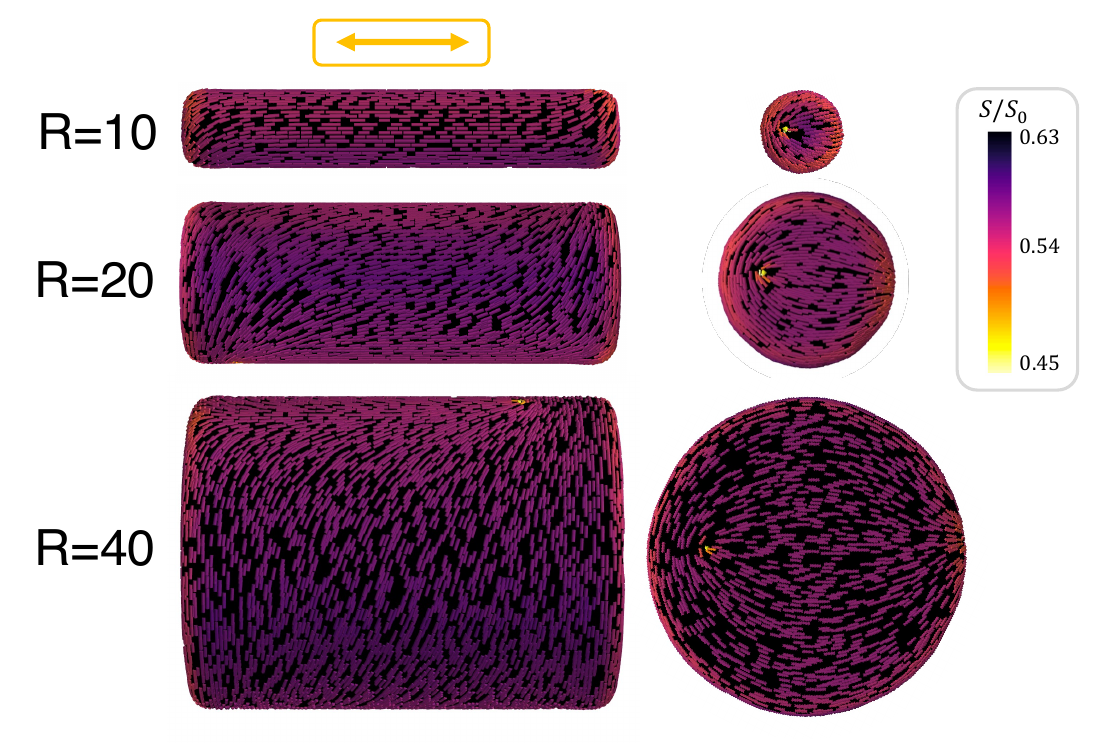}
\caption{\textbf{Director preferentially aligns with the cylinder long axis for small radii}. Left gives the side view of cylinder and right views the cylinder face. Orange arrow represents the geometry-induced alignment axis which has a field strength that scales with elastic constant $K$ and cylinder radius $R$ as $K/R^2$. Director coloured by scalar order parameter $S$ scaled by the equilibrium value $S_0$.
}
\label{fig:alignment}
\end{figure}
The simulations use strong planar degenerate anchoring to align the director field tangentially to the cylindrical surface. As a consequence of the Poincar\'{e}-Hopf theorem, a net surface topological charge is imposed that is equal to the Euler characteristic of the cylinder, $\sum_i k_i=\chi$. Here, $k_i$ is the winding number of each defect $i$ and the Euler characteristic of the cylindrical surface $\chi=2$ is the same as a sphere.
To satisfy these charge constraints, surface defects arise that tightly bind to the surface in a point-defect configuration known as a boojum \cite{mermin1990} or as line defects that smoothly connect between points on the surface. 
The latter are typically observed with stronger anchoring \cite{Liu2013}. 
In our equilibrium simulations, performed on cylinder radii between $R=10$ and $R=40$ and fixed length $L=108$ (simulation units), two short disclination lines are formed close to the edges of the cylinder and each connect between two $+1/2$ endpoints (\fig{fig:localiseddelocalised}a-c). 
These defects are elastically drawn to the edges in order to minimise the strong deformations imprinted by the cylindrical surface onto the bulk director. 
The coupling of defects and high surface curvature is well known in passive and active liquid crystals \cite{turner2010,tran2016,ellis2018} and forms the basis of a \textit{localised} state, where defects are locked to high-curvature regions of the confinement. 

\begin{figure*}[t]
\centering
\includegraphics[width=0.8\linewidth]{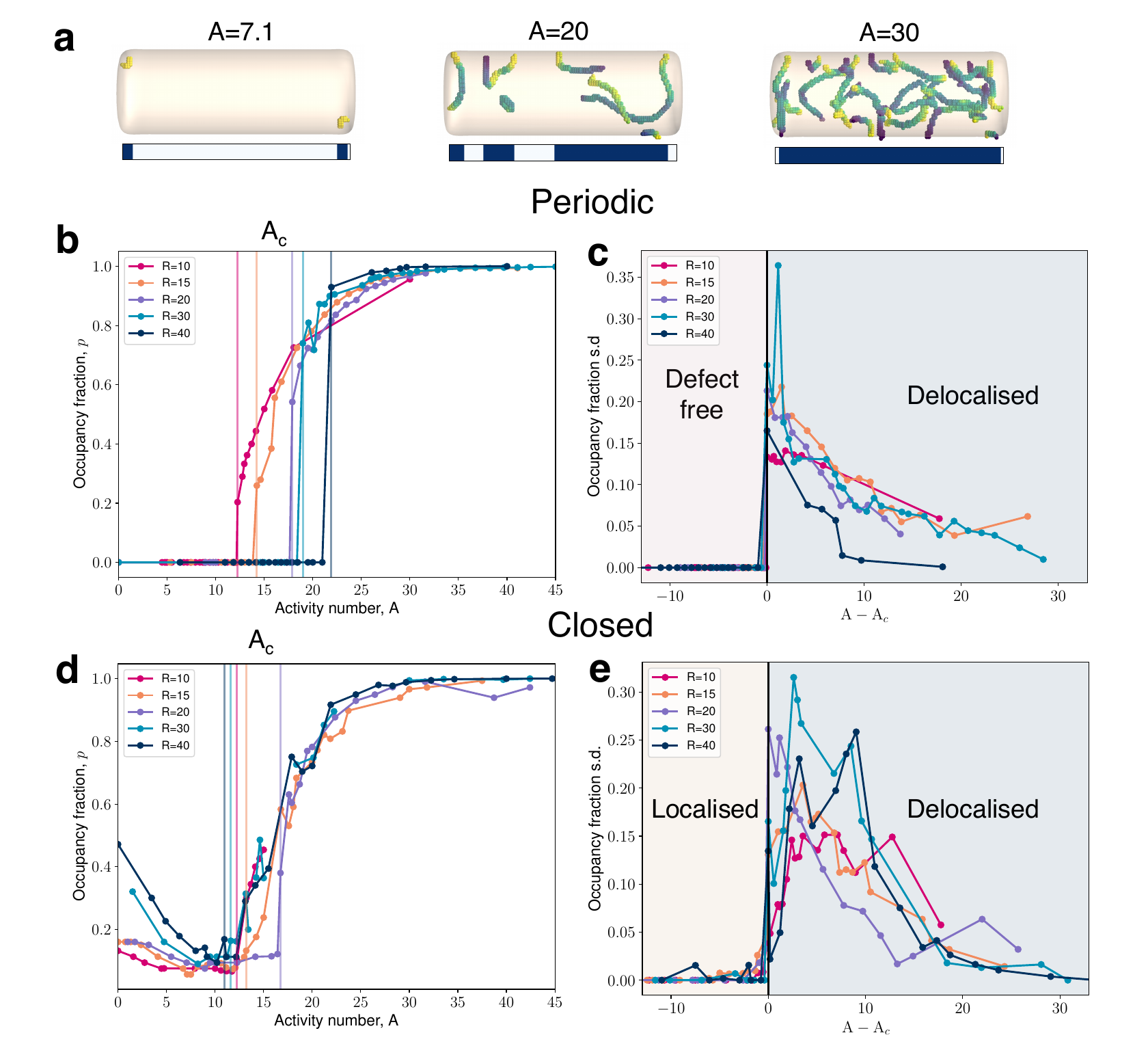}
\caption{\textbf{Characterisation of defect states via the occupancy fraction} $p$. \textbf{a}, Snapshots of closed cylinders and their defects for three different activity number values $\mathrm{A}$. Below each snapshot is a bar representing the fraction of the cylinder occupied by defects, where occupied regions are shown in dark blue and unoccupied regions in pale blue. 
\textbf{b}, Occupancy fraction as a function of dimensionless activity number $\mathrm{A}$ shown for different radii $R$. Delocalisation transition threshold $\mathrm{A_c}$ indicated as solid vertical lines with colour corresponding to $R$. \textbf{c}, Standard deviation in $p$ with distance from each radius' delocalisation threshold $\mathrm{A-A_c}$.
\textbf{d}, Same as \textbf{b} but for a closed cylinder. \textbf{e}, same as \textbf{c} but for a closed cylinder.}
\label{fig:orderparam}
\end{figure*}

Within the equilibrium localised state, the disclination line length increases with cylinder radius (\fig{fig:localiseddelocalised}a). 
This indicates another important consequence of the cylindrical geometry, in addition to edge-induced defect–curvature coupling, which is the influence of extrinsic curvature along the cylinder axis. The director favours alignment along the cylinder since azimuthal orientations cost elastic free energy via bend deformations. 
This curvature effect acts as a geometric field which aligns the director along the cylinder (\fig{fig:alignment}) and scales as $\sim K/R^2$, where $K$ is the elastic constant under the one-constant approximation \cite{stewart2004}. Consequently, defects localise closer towards the end of the cylinder when the radius is small, resulting in shorter disclination lines.

Defects may move freely throughout the cylinder to form a \textit{delocalised} state through the introduction of activity. 
Owing to strong deformation around their core, defects are dominant sources of active force which renders them motile and essential drivers of emergent flows \cite{giomi}. The snapshots in \fig{fig:localiseddelocalised}b-c highlight that delocalised states typically contain numerous defects which deform and rewire under chaotic active flows. 
Most of these defects are lines which connect several combinations of surface defects, either two $+1/2$, two $-1/2$, or a $+1/2$ and $-1/2$ defect, which collectively satisfy the $+2$ surface charge constraint. Infrequently, defect loops arise (Supplementary Movie 1). 
Both lines and loops exhibit a range of local defect profiles, characterised by the relative orientation of two vectors: the rotation vector $\mathbf{\Omega}$, which defines the winding plane of the director around the defect, and the tangent vector $\mathbf{T}$ \cite{friedel1969}.
The angle between these vectors is the twist angle $\beta$, which marks the local profile as $+1/2$ wedge when $\cos\beta=+1$, $-1/2$ wedge when $\cos\beta=-1$ and twist when $\cos\beta=0$. While equilibrium defects in the cylinder are pure $+1/2$ disclinations, activity results in defects displaying a wide range of winding patterns. The topological patterns in cylindrical confinement, and dependence on activity, are explored in the section \textit{Delocalised state}.

\begin{figure*}
\centering
\includegraphics[width=1.0\linewidth]{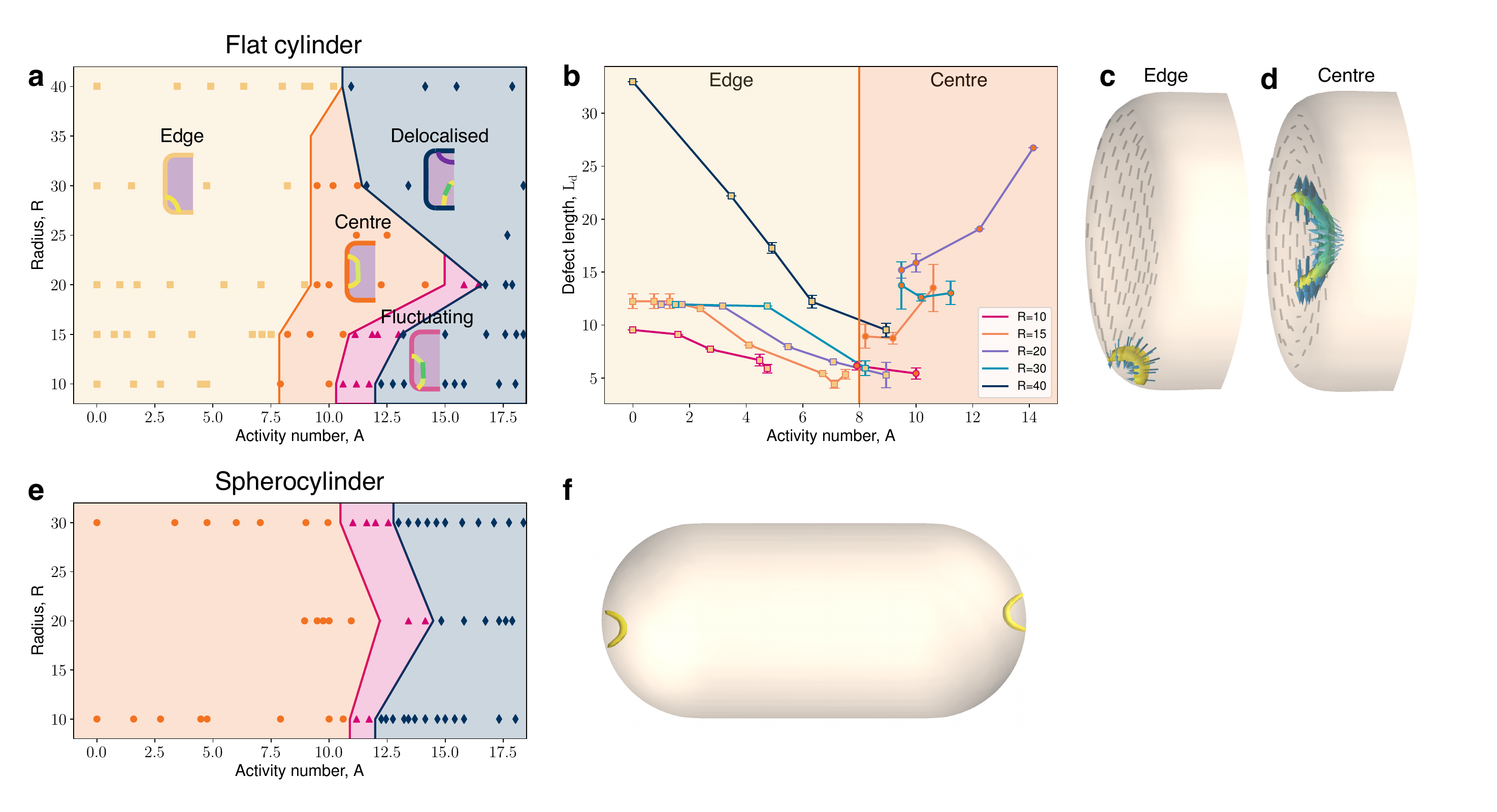}
\caption{\textbf{Activity-driven transitions within the localised state.} \textbf{a}, Regime map for defect localisation sites in terms of cylinder radius $R$ and activity number $\mathrm{A}$. Defects reside at cylinder edges (yellow shading and square markers), or at the centre of the cylinder face (orange shading and circle markers), or they fluctuate around the cylinder endcap (pink shading and triangle markers) when localised. Delocalised defect states are shown as blue shading and diamond markers. Transition thresholds between regimes are solid lines.
\textbf{b}, Defect contour length $L_d$ variation with activity, separated for edge (square markers; yellow shading) and centre (circle markers; orange shading) states. Error bars are standard deviation in $L_d$. \textbf{c,d,} Active force variation (blue arrows) along the defect in the edge state \textbf{c}, and centre state \textbf{d}. 
\textbf{e}, Regime map for defect states in spherocylinder geometries. \textbf{f}, Snapshot of defects in an $R=30$ spherocylinder at equilibrium ($\mathrm{A}=0$).
}
\label{fig:localised}
\end{figure*}

\subsection{Characterising defect states}

We next characterise the activity-driven transition between localised and delocalised defect states. 
The dimensionless activity number is employed as a control parameter, 
\begin{align}
    \textrm{A}=\sqrt{\frac{R^2 \zeta}{K}},
\end{align}
which is the ratio of the confining size, namely the cylinder radius $R$, to the active lengthscale of the active nematic $\ell_\zeta=\sqrt{K/\zeta}$. The active lengthscale is a characteristic scale set by the competition between active and elastic stresses \cite{hemingway2016}.
The activity number has had great utility in identifying the activity-driven transitions, such as the transition to active turbulence \cite{Doostmohammadi2017,shendruk2018} and regime maps for emergent flow states in confinement \cite{keogh2022,negro2025,carenza2019}.
To quantify the transition to delocalisation, we define an order parameter called the defect occupancy fraction
\begin{align}
    p=\frac{\langle N_\parallel\rangle_t}{L},
\end{align}
where $L$ is the length of the cylinder and $N_\parallel$ is the number of simulation cells in the axial direction that are occupied by a defect. The fraction of the cylinder occupied by defects is visually represented in \fig{fig:orderparam}a.
The average $\langle \cdots\rangle_t$ is performed over all simulation time $t$, excluding an initial warm-up period. 
Small $p$ is characteristic of a localised state, where the defects are small and confined to the endcaps, while $p=1$ indicates that the defects span the entire cylinder.


To explore the onset of delocalised defects we first consider a cylinder with periodic boundary conditions in the axial direction. Unlike the closed cylinder, this confining geometry is defect free in equilibrium, since it is topologically equivalent to a 2-torus for which $\chi=0$. We initialise the director field 
to be primarily aligned along the long axis of the cylinder but with a small perturbation, so that the instability can take hold at sufficient activity. When $\mathrm{A}$ is small, elasticity restores perfect alignment along the cylinder and $p=0$ (\fig{fig:orderparam}b).
Above a critical activity number $\mathrm{A}_c$, the perturbation grows with the active hydrodynamic instability \cite{simha2002} and defects proliferate within the channel. This onset of defects can be observed in \fig{fig:orderparam}b,c where $p$ and standard deviation of 
$p$ sharply jumps. 
Here, all defects are delocalised because there is no energetic preference to favour specific regions along the cylinder. 


In the closed cylinder (\fig{fig:orderparam}d), distinctive regimes can be observed where
$p$ decreases with activity (when $\mathrm{A}\lesssim12$) and increases with activity (when $\mathrm{A}\gtrsim12$).
The delocalisation threshold $\mathrm{A}_c\approx 12$ sits close to the minimum of $p$, in-between these regimes. This threshold 
is determined from visual inspection of delocalised defects within the cylinder, but equivalently corresponds to the activity value where the fluctuations in $p$ spike (\fig{fig:orderparam}e). 
For this sufficiently large activity, active stress around defects can exceed the elastic localisation barrier, leading to bursts of motility, stretching, and hence strong fluctuations in $p$. The spike in fluctuations of $p$ is also a strong indicator of delocalisation as a non-equilibrium phase transition.

\begin{figure*}
\centering
\includegraphics[width=1.0\linewidth]{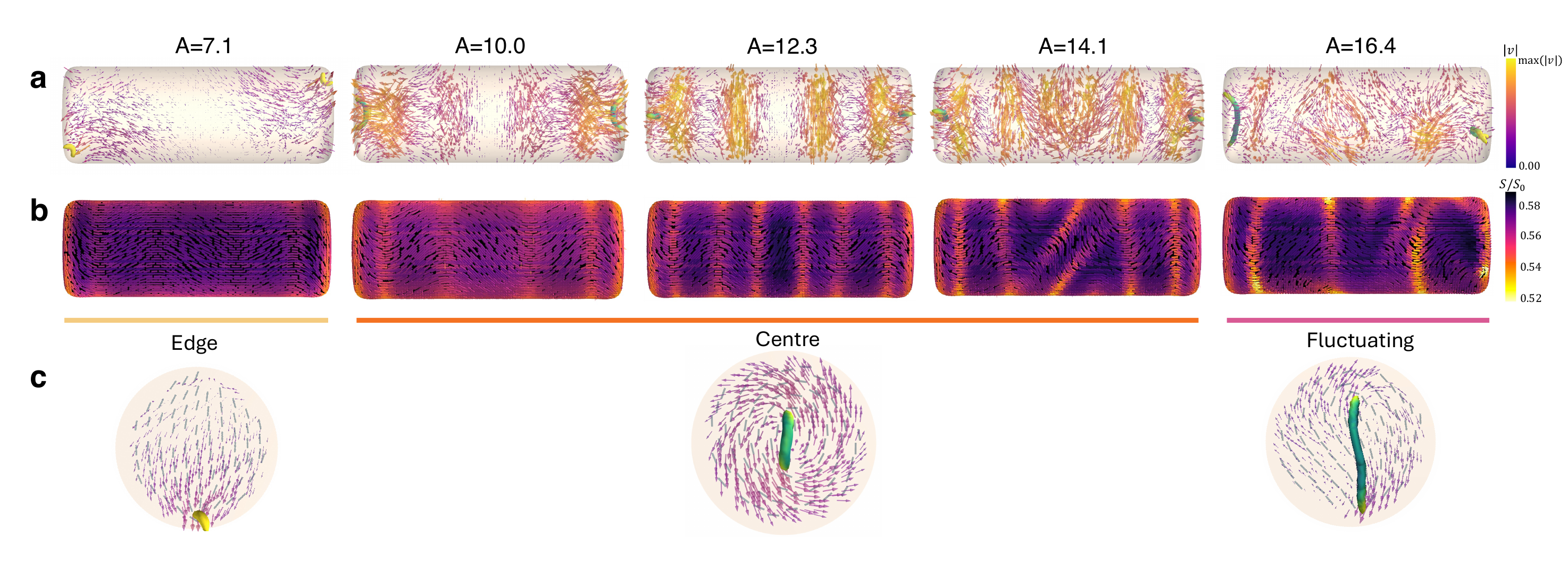}
\caption{\textbf{Flow and director field snapshots for different localisation states}.  \textbf{a}, Snapshots of the bulk velocity field for $R=20$. The snapshots are ordered by activity number $\mathrm{A}$. Velocity arrows are coloured by velocity magnitude $|v|$. \textbf{b}, Same as \textbf{a}, but for director field on the cylinder surface. Director coloured by scalar order parameter $S$ scaled by the equilibrium value $S_0$. \textbf{c}, Defects and flow field on the cylindrical face shown for $\mathrm{A}=7.1,10.0$ and $16.4$. Disclination lines are coloured by $\cos\beta$ (see \fig{fig:localiseddelocalised}).   
}
\label{fig:localisedflows}
\end{figure*}

For periodic and closed cylinders, the curves of $p(A)$ collapse for varying $R$ above the delocalisation threshold ($\mathrm{A}>\mathrm{A}_c$; \fig{fig:orderparam}b,d) and fully span the cylinder with $p\rightarrow 1$ at $\mathrm{A}\approx 30$. However, a striking difference between geometries is the threshold of $\mathrm{A}_c$. While $\mathrm{A}_c$ is comparable between closed cylinders, $\mathrm{A}_c$ increases with $R$ for periodic cylinders. 
This is an interesting observation for a couple of reasons. 
Firstly, differing thresholds indicate that there are alternative mechanisms that drive the delocalisation transition in closed cylinders. This is investigated in the section \textit{Delocalisation transition}. 
Secondly, it is the activity number alone that governs the threshold between regimes in confined active dynamics \cite{Doostmohammadi2017}, including the onset of spontaneous flows on cylindrical surfaces \cite{napoli2020}, yet here the radius dependence on $\mathrm{A}_c$ indicates that another lengthscale could be relevant to the transition to delocalisation in 3D cylinders. It would therefore be insightful to explore the radius dependence of the onset of defect formation in periodic cylinders in future studies.


In the sections to follow, we explore and characterise the rich defect behaviours in the regimes identified by $p(A)$ for closed cylinders. 

\subsection{Localised state}
\begin{figure}[t]
\centering
\includegraphics[width=0.85\linewidth]{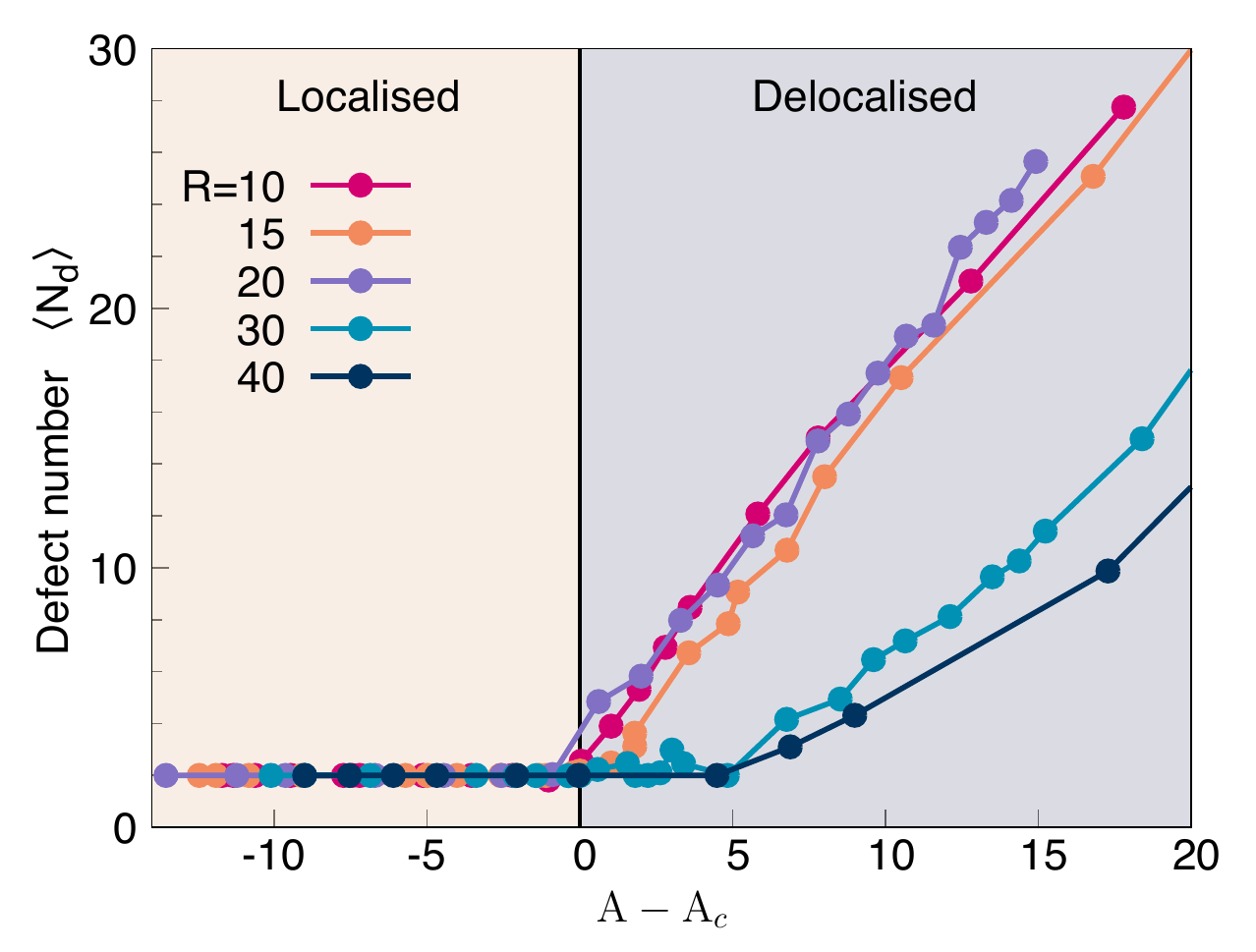}
\caption{\textbf{Defect number variation across the delocalisation transition.} Activity number $\mathrm{A}$ characterises distance from delocalisation threshold $\mathrm{A}_c$. Black vertical line indicates $\mathrm{A}=\mathrm{A}_c$,
which separates localised (yellow shading) and delocalised states (blue shading).}
\label{fig:numberdensity}
\end{figure}
Within the localised state, defects can be confined to the endcaps in several different ways. 
Defects can be locked at the cylinder edges (\fig{fig:localised}c; Supplementary Movie 2), reside at the centre of the cylinder face (\fig{fig:localised}d; Supplementary Movie 3), or fluctuate around the endcaps (Supplementary Movie 4).
In \fig{fig:localised}a, the three possible defect localisation states are presented in a regime map as a function of the cylinder radius and activity number. 
For small $\mathrm{A}$, defects reside at the cylinder edges, as in the equilibrium configurations (\fig{fig:localiseddelocalised}a). The centre and fluctuating regimes are reached at higher activities ($8\lesssim A \lesssim A_c$), with the latter observed only in a small region of parameter space for small radii ($R\leq20$) near the delocalisation threshold. 

How does activity influence defects within these regimes? In \fig{fig:localised}b we show the variation of defect lengths $L_d$ with activity number $A$. 
Within the edge state, increasing activity shortens the defect line, while in the centre state it has the opposite effect, stretching the defect line. 
In both states, the defect lengths are mostly constant in the steady state, as indicated by the small standard deviation of $L_d$ (see errorbars). 
The defect lengths are set by the balance of active and elastic forces. For the edge state, active forces are directed inwards towards the edge of the cylinder--towards the bend direction of the +1/2 profile orientation (\fig{fig:localised}c). Since both orientations are towards the same edge, this has a contracting effect on the disclination line. In contrast, the two $+1/2$ endpoints of the centre state, and associated active forces, are directed outwards from the centre (\fig{fig:localised}d). This forcing acts to extend the defect line as activity is increased.
These length variations identify why the defect occupancy fraction $p$ (\fig{fig:orderparam}d) has a decreasing trend, at small $\mathrm{A}$, when in the edge state. We note that the centre state is less clear in \fig{fig:orderparam}d as it exists only over a small range of $\mathrm{A}$ and because the stretching of the lines occurs transverse to the cylinder axis rather than along it.
\begin{figure*}[t!]
\centering
\includegraphics[width=0.85\linewidth]{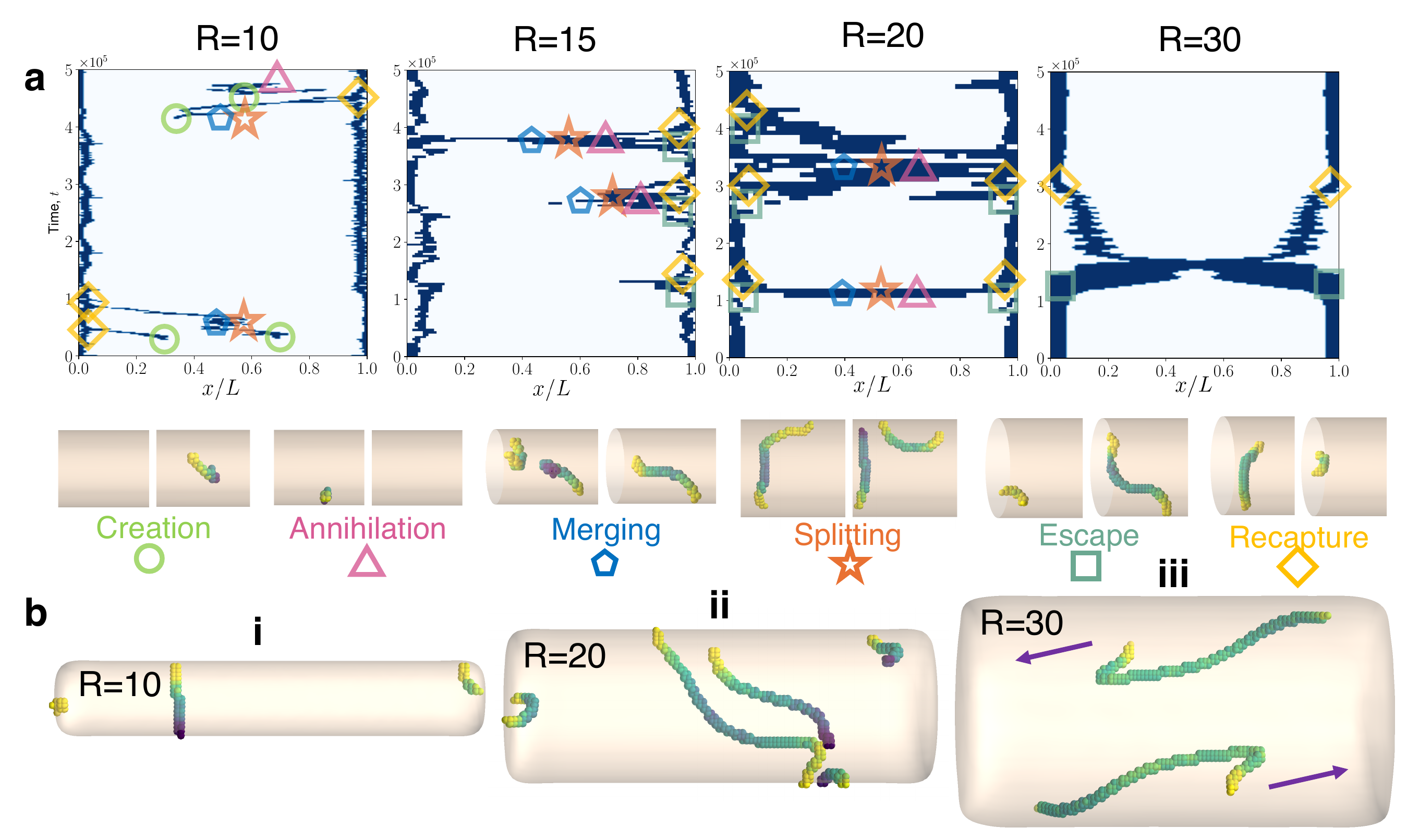}
\caption{\textbf{Defect dynamical processes at the delocalisation transition.} \textbf{a}, Kymographs of the defect occupancy along the cylinder of length $L$. Space in the axial direction, $x$, that is occupied by defects is shown in dark blue, while light blue is unoccupied. Symbols indicate the defect dynamical processes. 
New defects can be created (circle) or existing defects annihilated (triangle).
Two defect lines can join to form one line in a merging event (pentagon). A single defect line can form two defects in a splitting event (star). Defects can move away from the cylinder endcaps in an escape process (square) or be recaptured into a localised state (diamond). 
\textbf{b}, Snapshots of defects at the transition. \textbf{i}, Defects are created along the cylinder for $R=10$, with topologically-stable lines fixed at endcaps.  \textbf{ii}, Intermittent bursts of defects are characterised by defects escaping endcaps and undergoing merging, splitting and annihilation ($R=20$).  \textbf{iii}, Escaped defects in $R=30$ can smoothly move past each other, towards the opposite endcap, without touching. Defect direction of motion is indicated by purple arrows.}
\label{fig:occupancy}
\end{figure*}

We next investigate how activity facilitates transitions between localisation types. To explore this, we inspect the velocity fields that accompany the defect states (shown for $R=20$ in \fig{fig:localisedflows}). Flow structures in active nematics are correlated across scales comparable to the active lengthscale $\ell_\zeta$ \cite{Giomi2015}. 
Thus, at the smallest activities, the edge state displays the largest flow structures. These flows span the cylinder length and are directed towards the defects, consistent with the active force picture (\fig{fig:localised}c). 
As activities are raised to the centre-state threshold, the flows are correlated on smaller scales, leading to vortical flows that align along the length of the cylinder. 
One-dimensional arrays of vortices are a hallmark feature of pre-turbulent flow states in two-dimensional channels \cite{shendruk2017,hardouin2019} and three-dimensional square ducts \cite{Chandragiri2020, keogh2022}.
The vortices and their handedness originate from strong bend deformations (\fig{fig:localisedflows}b), which are greatest at the cylinder surface and drive shearing flows that revolve azimuthally around the cylinder. 
As activity is increased further, the spacing between bend walls reduces and more vortices fit along the channel. Eventually, the active lengthscale becomes smaller than the confinement size $R$, leading to vortices destabilising. This corresponds to the fluctuating regime where defects and flows depart from spatio-temporally ordered motion, but are still localised to the endcaps.

In addition to the balance between confinement $R$ and the characteristic size $\ell_\zeta$, the curvature of the cylinder ends also influences the localisation type (\fig{fig:localised}e). 
Since defects preferentially localise near the centre of curvature, the rounded caps of spherocylinders can cause defects to become fixed at the centre for $A\lesssim12$ (\fig{fig:localised}f), consequently removing the edge state from the regime map. This results in spherocylinders only having two possible localisation states: centre and fluctuating. Interestingly, the fluctuating state extends to a greater activity range and larger $R$ compared to flat cylinders. For flat cylinders, curvature is strongly concentrated at the edges which inhibits defects from moving within the endcap. In contrast, spherocylinders have uniform curvature throughout the endcap, which acts as a weaker but more spread out elastic potential that traps defects but permits movement within the endcaps, hence enhancing the fluctuating state.


\subsection{Delocalisation transition}
\label{sctn:delocalisation}

\begin{figure*}
\centering
\includegraphics[width=1.0\linewidth]{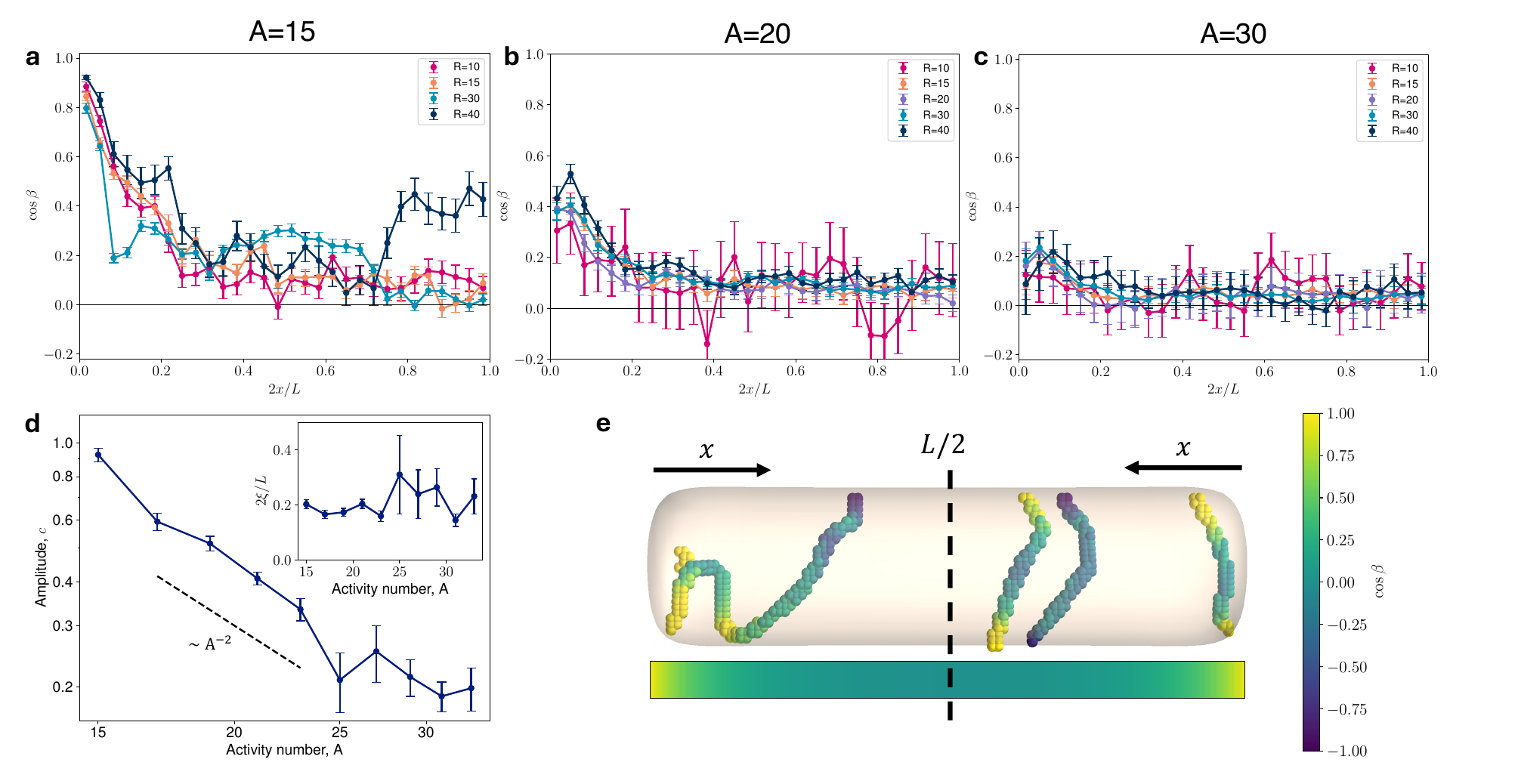}
\caption{
\textbf{Local charge segregation across the cylinder.} \textbf{a--c,} Mean $\cos\beta$ as a function of distance from the cylinder endcap for activity numbers $\mathrm{A}=15, 20,$ and $30$, respectively. Curves show different cylinder radii between $R=10$ and $R=40$. Error bars show standard error. 
\textbf{d,} Amplitude and localisation length (inset), shown as a function of activity number, obtained from exponential fits of \textbf{a--c}, merging $R$ values.
Amplitude scaling of $\mathrm{A}^{-2}$ shown as black dashed line. Error bars for figure and inset are standard deviation.
\textbf{e,} Snapshot of defect lines for $\mathrm{A}=15$ in a cylinder with $R=15$. Labels show distance $x$ measured from the cylinder ends, with $x = L/2$ at the cylinder centre. Coloured bar below shows exponential decay of $\cos\beta$ using the amplitude $c$ and localisation length $\xi$ for corresponding activity number in \textbf{d}. 
}
\label{fig:chargeseparation}
\end{figure*}

What are the mechanisms that lead to delocalisation? To explore this, we first examine the number of defects $N_d$ as the delocalisation transition is surpassed (\fig{fig:numberdensity}). 
Across all radii, $N_d$ is constant, consisting of only the two topologically-required disclinations, until an activity threshold is crossed and defects proliferate. This proliferation activity is exactly the delocalisation threshold for small radii $R\leq20$ but requires greater activity numbers within the delocalised state for $R>20$. 

To unpack how confinement controls defect behaviours at delocalisation, \fig{fig:occupancy}a shows defect occupancy kymographs for different radii at $\mathrm{A}=\mathrm{A}_c$. 
For $R=10$, a disclination line is confined to each endcap at all times.
At various intervals, a delocalised defect nucleates along the cylinder with 
a zero net surface charge due to its $+1/2$ and $-1/2$ wedge endpoints (\fig{fig:occupancy}b; left). 
These creation events occur at the surface of the cylinder (Supplementary Movie 5) where deformation and shearing flows are strong. This mechanism of surface defect creation resembles the hydrodynamic instability of extensile active nematic films \cite{martinez2019}, where $\pm 1/2$ defect pairs unbind along narrow walls of bend \cite{shankar_defect_2018,l_head}.
Following creation, defects are motile around the cylinder and transiently detach into defect loops.
The defects then proceed to self-annihilate, due to their charge neutrality, or merge with the topologically-stable lines at the endcaps. For $R=15$ and $R=20$, the kymographs reveal that defects reside at the endcaps for a portion of the simulation. 
However, eventually active fluctuations endow defects with sufficient motility to exceed the elastic localisation barrier and escape confinement (Supplementary Movie 6 and 7). 
Escaped defects undergo merging, splitting and annihilation (\fig{fig:occupancy}b; middle) before returning to the cylinder endcaps for another period of localisation. These features are seen as intermittent bursts of defects in the kymographs and are responsible for the increase in defect-count in \fig{fig:numberdensity} in $\mathrm{A}=\mathrm{A}_c$. 
Finally, for $R=30$, defects, too, overcome their confinement barrier and delocalise via escape. 
This follows from the expanding spiral trajectories traced by defects on the cylinder face as the vortices destabilise (Supplementary Movie 8). 
Following escape, defect motility drives the disclination lines to cross past each other before being recaptured by the opposite endcap for another period of localisation.  
These $R=30$ defects have sufficient space to pass each other without interacting (\fig{fig:occupancy}b; right), and consequently no additional defects are generated during delocalisation. This is why $R\geq30$ generates defects at activities values above delocalisation ($\mathrm{A}>\mathrm{A}_c$), while $R=15$ and $20$, which also delocalise via escape, form additional defects exactly at the delocalisation threshold ($\mathrm{A}=\mathrm{A}_c$) in \fig{fig:numberdensity}.

\subsection{Delocalised state}
\label{section:delocalised}

Delocalised defects exhibit markedly different behaviour to localised defects.  
In this high-activity regime, defects undergo chaotic dynamics which are characterised by continual resizing and rewiring that lead to variations in their number.
Here, we explore how elasticity and activity compete within a confined setting to control the local winding patterns of defects and the statistical properties of their number and length.

\subsubsection{Charge effects}

The local winding character is strongly linked to the localisation and motility of defects, which are controlled by passive elasticity and active driving respectively \cite{binysh2020}. 
As aforementioned, in the passive limit ($\mathrm{A}\ll \mathrm{A}_c$), the curvature couples with the defect to enforce $\cos\beta=1$ everywhere along the line (\fig{fig:localiseddelocalised}a). 
However, at higher activity numbers, this geometry effect competes with active forcing which introduces a wider range of winding characters (\fig{fig:localiseddelocalised}b,c).
In \fig{fig:chargeseparation}a-c, we show the variation of $\cos\beta$ along the cylinder (\fig{fig:chargeseparation}e) for three chosen activity numbers within the delocalised state, $\mathrm{A}=15, 20$ and $30$. 
At activities just above the delocalisation transition ($\mathrm{A}=15$), 
$\cos\beta\approx0.9$ at the cylinder endcaps, consistent with the elastic expectations. Away from the endcaps, $\cos\beta$ decays exponentially towards $\cos\beta\approx0$. We do not show $R=20$ in \fig{fig:chargeseparation}a as the delocalisation transition takes place at a higher activity ($\mathrm{A=16.7}$). Deeper into the delocalised phase, at $A=20$ and $A=30$, the endcaps exhibit $\cos\beta\approx 0.4$ and $\cos\beta\approx 0.2$, respectively, with both values decaying toward the centre. Significantly, for each activity number, $\cos\beta$ collapses for all cylinder radii. 
To understand how activity disrupts the curvature-induced influence on $\cos\beta$, we fit the form $c\exp(-x/\xi)$ to extract the amplitude $c$ and characteristic decay length $\xi$. 
Each fit of $\cos\beta$ includes all radii within a small interval of activity numbers, using a bin size of $2$.
The amplitudes and characteristic lengths are shown in \fig{fig:chargeseparation}d. 
Interestingly, $\xi$ is approximately constant across all activity numbers, while $c$ strongly depends on activity number, decaying with a power law of $\mathrm{A}^{-2}$. This indicates that activity directly tunes the strength of the geometry-induced winding, but does not influence the range over which the defects can interact with the confining geometry.


\subsubsection{Number of defects}

\begin{figure}[t]
\includegraphics[width=0.85\linewidth]{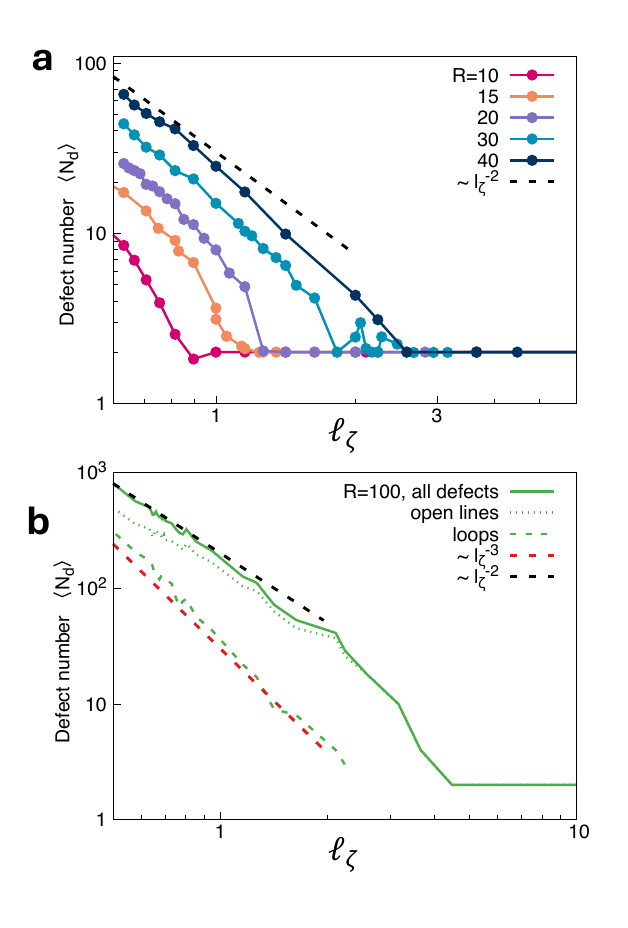}
\caption{\textbf{Average number of defects as a function of the active lengthscale}. \textbf{a,} Defect number for cylinder radii between $R=8$ and $R=40$ are solid lines. Corresponding scaling behaviour $\sim\ell_\zeta^{-2}$ shown as dashed black line. \textbf{b}, Average defect number with $R=100$ separated for defect lines (dotted), defect loops (dashed) and combined defects (solid). Scaling behaviours of $\langle N_d \rangle\sim \ell_\zeta^{-2}$, for defect lines, and $\langle N_d \rangle\sim \ell_\zeta^{-3}$ for defect loops, are dashed black and red lines respectively. 
}
\label{fig:numberscaling}
\end{figure}

We next investigate how activity influences the number of defects in the delocalised state. The scaling of the number of defects with activity has been characterised in bulk \cite{Digregorio2024}, but not within confined settings where disclination lines dominate. 
In \fig{fig:numberscaling} we show the variation of the average number of defects $\langle N_d\rangle$ with active lengthscale $\ell_\zeta$. 
After the onset of proliferation, the number of defects follows the scaling $\langle N_d\rangle \sim \ell_\zeta^{-2}$. This scaling is observed for $R\leq 40$ (\fig{fig:numberscaling}a) and for defect lines that connect to the boundaries in weak confinement $R=100$ (\fig{fig:numberscaling}b). The same scaling for both confinement types, demonstrates that this scaling is universal for defect lines in activity-dominate regimes.

In weak confinement, a subset of the defects are loops, which scale as $\langle N_d\rangle \sim \ell_\zeta^{-3}$, which is the same scaling as disclination loops in bulk active turbulence \cite{Digregorio2024}.  To understand why defect lines and loops scale differently, we turn to logic from two dimensions.
It is well characterised that the distance between defects scales with the active lengthscale \cite{hemingway2016}. As a consequence, $N_d\sim A_{2D}/\ell_\zeta^2$, where $A_{2D}$ is the system area. In our system, defect lines scale equivalently because they are constrained to the cylinder surface.  
This surface constraint thus enforces that $N_d\sim A_{cylinder}/\ell_\zeta^2$, where $A_{cylinder}$ is the cylinder surface area. In contrast, for defect loops, which reside entirely in the cylindrical bulk, the number of defects scale inversely with the active volume element as $N_d\sim \ell_\zeta^{-3}$.

\subsubsection{Contour lengths}

\begin{figure*}
\includegraphics[width=1.\linewidth]{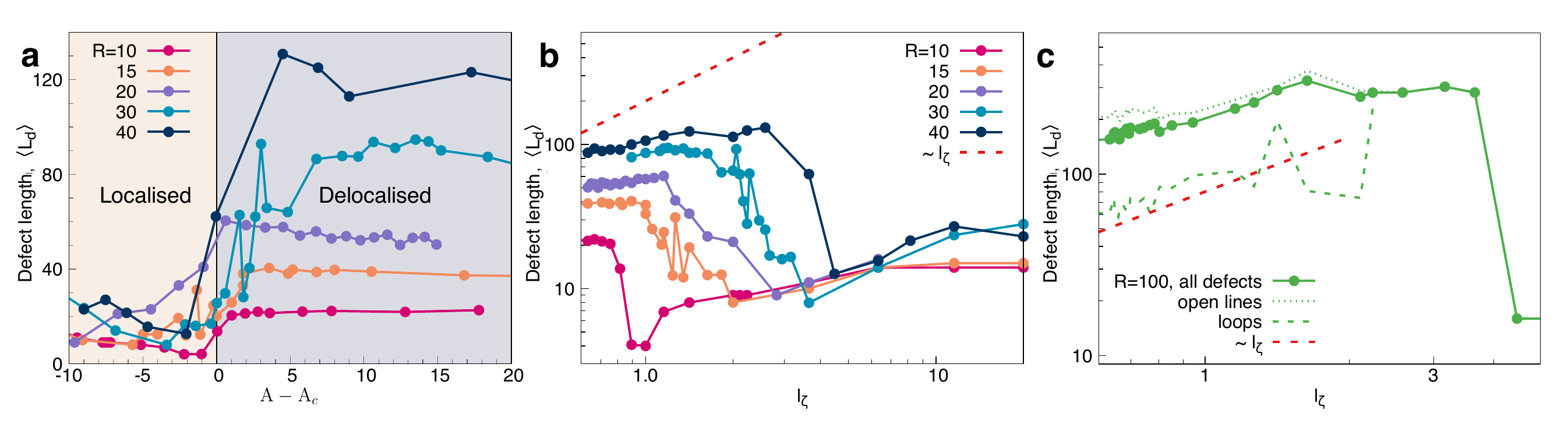}
\caption{\textbf{Average defect length as activity is varied.} \textbf{a,} Contour length as a function of activity number deviation from delocalisation threshold, $\mathrm{A}-\mathrm{A}_c$. Curves show different radii sampled between $R=10$ and $R=40$. 
\textbf{b}, Same as \textbf{a} but in terms of active lengthscale $\ell_\zeta$. \textbf{c}, Contour length for $R=100$ with defect lines (dotted), loops (dashed) and combined defects (solid) separated. Scaling of $\sim \ell_\zeta$ is indicated by dashed red line. }
\label{fig:scaling1}
\end{figure*}

\begin{figure*}
\includegraphics[width=1\linewidth]{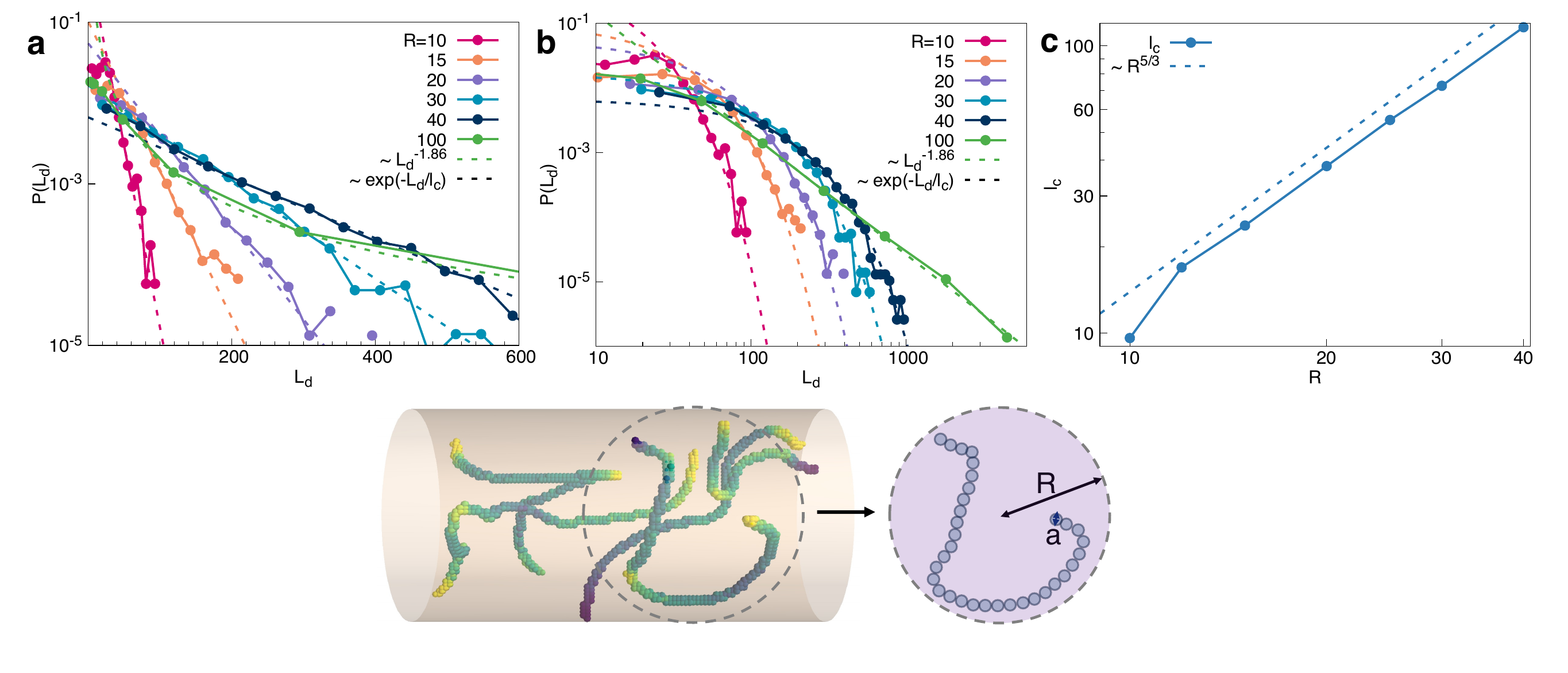}
\caption{\textbf{Defect length distributions and characteristic sizes in the delocalised state.} \textbf{a}, Probability density function of defect lengths, $L_d$, for different values of cylinder radii $R$. Exponential fits $\sim \exp{(-L_d/l_c)}$, with characteristic length $l_c$, shown by dashed straight lines on semi-log scale. \textbf{b}, Same as \textbf{a}, but now using a log-log scale. Power law scaling of $R=100$ indicated by green dashed line. \textbf{c}, Characteristic defect lengths as a function of cylinder radius. Power law scaling of $l_c\sim R^{5/3}$ shown by blue dashed line. \textbf{d}, Schematic of a defect conformation within a blob of radius $R$ which is discretised into monomers of size $a$.}
\label{fig:scaling2}
\end{figure*}

Activity has a significant influence on the lengths of three-dimensional defects, which has been primarily investigated for systems dominated by defect loops. The coarsening dynamics and steady-state length statistics are governed by a balance between active self-propulsion, set by the orientation and winding character of loop segments \cite{binysh2020}, and elastic line tension, which penalises small defect loops of high curvature \cite{kralj2023}. 
This competition establishes a linear scaling of defect length with the active lengthscale ($L_d\sim\ell_\zeta$) in bulk active turbulence \cite{Digregorio2024}, while topologically stable defect loops, studied in droplets, grow in average size with increasing activity (decreasing $\ell_\zeta$) \cite{copar2019, negro2025}. Here, we investigate how confinement impacts the contour length scaling of defect lines in activity-driven regimes. 

In \fig{fig:scaling1}a-c we show the variation of the mean defect length $\langle L_d\rangle$ with activity. As activity surpasses $\mathrm{A}_c$, defect lengths limit at a maximum size (\fig{fig:scaling1}a). This effect is most evident for strong confinement ($R\leq 20$), where the maximum size is reached at $\mathrm{A\approx A}_c$. 
At larger radii $R\geq 40$, defects reach a peak size at activity values above $\mathrm{A}_c$ beyond which the average defect length reduces with increasing activity (\fig{fig:scaling1}b). However, the activity-driven reduction in length for $R=40$ is less strong than the bulk $L_d\sim \ell_\zeta$ scaling. 
To test the limit towards bulk behaviour, we consider a cylinder radius of $R=100$ (\fig{fig:scaling1}c). 
For this larger radius, defects decrease in size with increasing activity and obtain the bulk scaling.

How does confinement limit the length of defects?  To explore restrictions on defect lengths we obtain the probability distribution of lengths $P(L_d)$ (\fig{fig:scaling2}a). 
The most probable defect lengths are $\approx R$, with an exponentially decaying tail for large defects, and the longest defects reaching $\approx 20R$. 
In comparison, weak confinement ($R=100$) shows heavy tails obeying a power law distribution (\fig{fig:scaling2}b) which aligns with the power law behaviour observed in the bulk limit \cite{Digregorio2024}. 
The reason for the suppressed tails in confinement is that large defects are strongly inhibited by the boundary, disconnecting into two or more lines when meeting the surface.

Returning to moderate confinement ($R\leq40$), we determine the characteristic size of  defects $\ell_c$ by fitting an exponential $P(L_d)\propto e^{-L_d/\ell_c}$. The characteristic size is set by the confinement size $R$, which can be observed in \fig{fig:scaling2}c where a power law emerges $\ell_c\sim R^{5/3}$.   
This power law can be interpreted through an analogy with the de Gennes blob theory of polymers in confinement \cite{micheletti2011}.
We discretise the disclination line into $N_g$ monomers of size $a$ arranged inside a `blob' of size $R$
\begin{align}
    R\sim aN_g^\nu.
\end{align}
With this discretisation, $a$ corresponds to the defect core size. The Flory exponent $\nu$ characterises the conformation of the defect, where $\nu=1/2$ is a simple random walk and $\nu=1$ is a fully stretched out chain.  
Since defects disconnect at boundaries, we assume that a single blob contains the disclination line. With this assumption, the contour length $l_c=aN_g$.  This gives the scaling
\begin{align}
    l_c \sim a^{1-\frac{1}{\nu}}R^{1/\nu}.
    \label{eq:lcinconfinement}
\end{align}
Therefore, this scaling analysis, together with \fig{fig:scaling2}c, identifies that $\nu=3/5$ which corresponds to the same exponent as a self-avoiding random walk.  
A self-avoiding random walk is reasonable since defects that exist in a given snapshot cannot self-intersect, else they would split into two shorter lines. Further, while $\nu=3/5$ is typical for polymeric systems with excluded volume interactions, here we have hydrodynamic interactions between the effective monomers.  It is plausible that the strong self-interactions through active forces drive defects into more stretched out configurations than a random walk.

\section{Conclusion}
In this work, we investigated the activity-tuned regimes of 3D active nematics in cylindrical confinement. Closed cylindrical geometries offer multiple utilities: they stabilise defect lines via surface topological charge, control defects through localised curvature at the cylinder edges, and establish a hydrodynamic screening length (cylinder radius) that is uniform along the channel. The ratio of the radius to the active lengthscale defines an important dimensionless number for the system, the activity number. Our results show that the activity number facilitates direct control of the defect pinning sites and vortex formation within the localisation regime, and the winding properties in the delocalised state.
\textcolor{black}{The transition from localised to delocalised defect configurations bears an intriguing qualitative resemblance to the transition between topological insulators and topological superconductors~\cite{bernevig2013,marchetti1993translational,balents1993delocalization,teo2010topological}, where in that context flux lines, rather than disclinations, either remain pinned to the surface or "escape" into the bulk.} 

\textcolor{black}{Our} results also highlight the similarities and differences between defect scaling properties in confinement vs bulk active turbulence. 
This work complements findings that defect loops scale as $N\sim\ell_\zeta^{-3}$ and uncovers that defect lines scale equivalently to two-dimensional defects $N\sim\ell_\zeta^{-2}$ since they end on the cylinder surface. 
Defect contour lengths in weak confinement scale like the bulk system $L_D\sim\ell_\zeta$, but confinement induces an activity-independent characteristic size that scales with the cylinder radius as $\sim R^{5/3}$. This analysis identifies that defect lines in cylinders behave as polymers in confinement with a self-avoiding random walk conformation. It would be valuable to deeper explore analogies between defects and polymers and this would further observations that 3D defects have qualitatively similar stretching and writhing dynamics to living active polymers \cite{cates1987}.

This work poses immediate extensions for active microfluidics in cylindrical capillaries: where one could tune the activity number and utilise surface topological constraints, via chambers, to induce spontaneous flows, place vortices and control defects in accordance with surface geometry. With the emerging biological relevance of three-dimensional active nematics, such as in-vivo gliomas \cite{argento_gliomas_2025}, mechanisms for localising defects may be insightful in controlling apoptosis and cell aggregate morphologies.





\section{Methods}

\subsection{Simulation details}
Here we outline the hydrodynamic model used in this work. We describe the physics of a liquid crystal double emulsion in terms of the following coarse-grained quantities: (i) the global fluid velocity ${\bf v}({\bf r}, t)$, (ii) a set of passive scalar phase field  $\phi({\bf r}, t)$, capturing the cylindrical confinement  (iii) a tensor order parameter ${\bf Q}({\bf r},t)$ accounting for the ordering properties of a liquid crystal made of rod-like molecules. In the uniaxial approximation, $Q_{\alpha\beta}=q(n_{\alpha}n_{\beta}-\frac{1}{3}\delta_{\alpha\beta})$ (Greek subscripts denote Cartesian components), where ${\bf n}$ represents the local orientation of the liquid crystal molecules (often termed director)
and $q$ gauges the amount of local order, a quantity proportional to the
largest eigenvalue of ${\bf Q}$ ($0\leq q\leq 2/3$).

The ground state of the system is encoded in the following free energy density
\textcolor{black}{\begin{eqnarray}\label{free}
&f&= \frac{a}{4}\phi^2(\phi-\phi_0)^2 + \frac{k}{2}(\nabla\phi)^2  \nonumber
\\&&+\frac{A_0}{2}\left(1-\frac{\chi(\phi)}{3}\right)Q_{\alpha\beta}^2-\frac{A_0\chi(\phi)}{3}Q_{\alpha\beta}Q_{\beta\gamma}Q_{\gamma\alpha}\nonumber\\&&+\frac{A_0\chi(\phi)}{4}(Q^2_{\alpha\beta})^2+\frac{\kappa}{2}(\partial_{\gamma}Q_{\alpha\beta})^2\nonumber\\&&+W\left[\partial_{\alpha}\phi Q_{\alpha\beta}\partial_{\beta}\phi\right]. 
\end{eqnarray}}
Here the first term, multiplied by the positive constant $a$, represents a double-well potential that ensures the existence of two coexisting minima at $\phi=\phi_0$, corresponding to the interior of the cylinder, and $\phi=0$ outside. 
The second term in Eq.(\ref{free}), multiplied by the elastic constant $k$, controls the interfacial energy. 
Both constants $a$ and $k$ fix
the surface tension $\sigma=\sqrt{{8ak/9}}$ and the interface thickness $\xi_{\phi}=\sqrt{2k/a}$ of the cylinder.
In all simulations we used $a=0.01$ and $k=0.14$.
The bulk properties of the liquid crystal are described by a fourth order expansion of the Q-tensor
(summation over repeated indices is assumed), where $A_0=0.5$ is a positive constant and $\chi(\phi)$, controls the isotropic-liquid crystal transition, which occurs when $\chi(\phi)>\chi_{cr}=2.7$ \cite{degennes}.
Following previous works \cite{sulaiman}, we set $\chi=\chi_0+\chi_s\phi$, where  $\chi_0$ and $\chi_s$ control the boundary of the coexistence
region. Note that $\chi$ depends exclusively on $\phi$, since the liquid crystal is confined solely within the layer where $\phi=\phi_0$. 
We fixed $\chi_0=2.4$ and $\chi_s=0.4$, so that, $\chi=3.2$ inside the cylinder. This gives an equilibrium value for the scalar order parameter $q=0.556$, in absence of activity. 
The energetic cost due to liquid crystal distortions is gauged by the gradients of $Q_{\alpha\beta}$ (where $\kappa=0.04$ is the elastic constant) while the anchoring 
of the director at the droplet interface is described by the last term where $W=0.2$ is the anchoring strength. The anchoring is tangential when $W>0$, the case considered here, otherwise it is perpendicular.


The dynamics of the scalar fields $\phi$  obeys a Cahn-Hilliard equation
\begin{equation}\label{cahn_eqn}
\partial_t \phi+{\bf v}\cdot{\nabla\phi}=M\nabla^2\mu,
\end{equation}
where $M$ is the mobility (set to $M=0.1$) and $\mu=\frac{\delta{\cal F}}{\delta\phi}$
is the chemical potential, with ${\cal F}=\int_V f d{\bf r}$.

The time evolution of the ${\bf Q}$ tensor is governed by the Beris-Edwards equation \cite{beris}
\begin{equation}\label{beris_eqn}
(\partial_t + {\bf v}\cdot\nabla){\bf Q}-{\bf S}({\bf W},{\bf Q})=\Gamma{\bf H},
\end{equation}
where $\Gamma$ is a collective rotational diffusion constant (set to $\Gamma=0.3$) and ${\bf H}$ is the molecular field, which is given by
\begin{equation}
{\bf H}=-\frac{\delta {\cal F}}{\delta {\bf Q}}+({\bf I}/3)Tr\frac{\delta {\cal F}}{\delta {\bf Q}},
\end{equation}
with ${\bf I}$ unit matrix. 
The first term on the left-hand side of Eq.\ref{beris_eqn} represents the material derivative accounting for the time dependence of a quantity advected by the fluid velocity ${\bf v}$. The second term is further contribution accounting for rotation and  stretching  of the rod-like molecules of the liquid crystals due to flow gradients \cite{beris}, and is given by
\begin{eqnarray}
S({\bf W},{\bf Q})&=&(\xi{\bf D}+{\bm\omega})({\bf Q}+{\bf I}/3)+
(\xi{\bf D}-{\bm\omega})({\bf Q}+{\bf I}/3)\nonumber\\&&
-2\xi({\bf Q}+{\bf I}/3)Tr({\bf Q}{\bf W}).
\end{eqnarray}
Here $Tr$ denotes the tensorial trace, while ${\bf D}=({\bf W}+{\bf W}^T)/2$ and ${\bm\omega}=({\bf W}-{\bf W}^T)/2$ are the symmetric and anti-symmetric part of the velocity gradient tensor $W_{\alpha\beta}=\partial_{\beta}v_{\alpha}$. The role of the constant $\xi$ is twofold. On the one hand, 
it determines the aspect ratio of the liquid crystal molecules; it is  positive for rod-shaped molecules and negative for disk-like ones. On the other hand, it 
controls  whether the director field is flow aligning or flow tumbling under shear. In such conditions, at the steady state the director would align with the flow gradient at
an angle $\theta$ such that $\xi cos(2\theta) = (3q)/(2 + q)$ \cite{degennes}, whose real solutions (corresponding to a flow aligning regime) are obtained if $\xi\geq 0.6$. In our simulations we fix $\xi=0.7$. 

Finally, the fluid velocity ${\bf v}$ is ruled by the incompressible Navier-Stokes equations
\begin{equation}\label{cont_eqn}
\nabla\cdot{\bf v}=0,
\end{equation}
\begin{equation}\label{nav_stok_eqn}
\rho(\partial_t{\bf v}+{\bf v}\cdot \nabla{\bf v})=-\nabla p + \nabla\cdot ({\bm \sigma}^{visc}+{\bm \sigma}^{lc}+{\bm \sigma}^{int}+{\bm \sigma}^{act}),
\end{equation}
where $\rho$ is the fluid density and $p$ is the hydrodynamic pressure. The viscous contribution is given by
\begin{equation}
\sigma_{\alpha\beta}^{visc}=\eta(\partial_{\alpha}v_{\beta}+\partial_{\beta}v_{\alpha}),
\end{equation}
where $\eta$ is the shear viscosity of the fluid, while the elastic one due to liquid crystal deformations reads
\begin{eqnarray}
\sigma_{\alpha\beta}^{lc}=&&-\xi H_{\alpha\gamma}(Q_{\gamma\beta}+\frac{1}{3}\delta_{\gamma\beta})-\xi(Q_{\alpha\gamma}+\frac{1}{3}\delta_{\alpha\gamma})H_{\gamma\beta}\nonumber\\
&&+2\xi(Q_{\alpha\beta}-\frac{1}{3}\delta_{\alpha\beta})Q_{\gamma\mu}H_{\gamma\mu}+Q_{\alpha\gamma}H_{\gamma\beta}-H_{\alpha\gamma}Q_{\gamma\beta}\nonumber\\
&&-\partial_{\alpha}Q_{\gamma\mu}\frac{\partial f}{\partial(\partial_{\beta}Q_{\gamma\mu})}.
\end{eqnarray}
A further contribution stems from interfacial stress between the active and the passive phase, and reads as follows,
\begin{equation}
\sigma^{int}_{\alpha\beta}=\left[\left(f-\phi\frac{\delta{\cal F}}{\delta\phi}\right)\delta_{\alpha\beta}-\frac{\partial f}{\partial(\partial_{\beta}\phi)}\partial_{\alpha}\phi\right].
\end{equation}
The last term is the active stress given by \cite{marchetti,hatwalne}
\begin{equation}
\sigma_{\alpha\beta}^{act}=-\zeta Q_{\alpha\beta},
\end{equation}
where $\zeta$ is the active parameter, positive for extensile materials and negative for contractile ones.

As in previous works \cite{tjhung2015,negro2025}, we use a 3D hybrid lattice Boltzmann method, which solves Eq.\ref{cahn_eqn} and Eq.\ref{beris_eqn} via a finite difference scheme while Eqs.\ref{cont_eqn} and Eq.\ref{nav_stok_eqn} by a standard LB approach. 

In all our numerical experiments, the Reynolds number is defined as $
Re = \dfrac{\rho \bar{v} R}{\eta},$
where $\rho = 2$ is the density of the incompressible fluid, $R=10-100$  is the confinement scale (i.e., the radius of the cylinder), $\eta = 1.7$  is the viscosity of the lattice Boltzmann algorithm, and $\bar{v}$ is a reference scale for the strength of the active flow. In our simulations it never exceeds $0.01$,  such that in all cases $Re < \mathcal{O}(1)$.\\
It is also worth mentioning that in our setup the viscosity of the embedding fluid and that of the active nematic are the same, as commonly assumed in lattice Boltzmann simulations of active gel hydrodynamics \cite{tjhung2015,ruske2021}.
The reasoning behind this choice is founded on the observation that, in experiments, active nematics are typically confined at the interface between water and oil, where the viscosity of water is of the same order of magnitude as the shear viscosity of active nematics, while the oil one  can be tuned across a wide range, from \(\eta \approx 10^{-3} \text{ Pa} \cdot \text{s}\)  to \(\eta \approx 1 \text{ Pa} \cdot \text{s}\), depending on the type of oil used \cite{Guillamat2016}. 
We expect that variations in the viscosity of the fluid within the cylinder, or that of the embedding fluid relative to the viscosity of the bulk active fluid, would lead to quantitative changes in the results presented in this article (e.g., the transition point of the delocalisation transition), whilst the qualitative features would remain essentially unchanged. 
The system is initialised with the phase field $\phi$ set to a cylindrical profile, with $\phi=\phi_0=2$ inside the cylinder and $\phi=0$ outside, and with the nematic tensor ${\bf Q}=0$ everywhere. The phase field is then evolved for $10^4$ time steps in the presence of hydrodynamics, to relax the interface. After this stage, the evolution of $\phi$ is frozen, and the ${\bf Q}$-tensor is initialised from a random configuration and let evolve towards its free-energy minimum. The active stress is switched on only on equilibrated passive configurations.


\subsection{Defect analysis}

Analysis of defect lines utilise the disclination density tensor proposed by Schimmings and Vi\~nals \cite{schimming2022}
\begin{align}    D_{ij}=\epsilon_{i\mu\nu}\epsilon_{jlk}\partial_l Q_{\mu\alpha}\partial_k Q_{\nu\alpha},
\end{align}
where $Q$ is the nematic tensor order parameter and $i,j,k,\alpha,\mu,\nu$ are tensor indices with applied summation convention. The tensor has the convenient representation as the dyad composing of the local line tangent $\mathbf{T}$ and the rotation vector $\mathbf{\Omega}$
\begin{align}
    \label{eq:Dtensordecomposition}
    D_{ij} = s(\mathbf{r})\Omega_{i}T_j,
\end{align}
where the scalar field $s(\mathbf{r})$ is positive and maximum at the disclination core. In this work, defects are identified as isosurfaces where $s(\mathbf{r})=0.09$ and the winding character, $\cos\beta$, is obtained from the trace of the disclination tensor. Visualisations of defects use the Mayavi library \cite{ramachandran2011mayavi}.

For extracting the defect number and contour lengths, we use the methods described in \cite{Digregorio2024}. 
Defect points are identified by the rotation of the nematic field, while looping over the contour of the faces of each 3D voxel in the simulation lattice. 
Faces that are concatenated with a disclination line correspond to a rotation $\pi$ of the nematic field, while the defect free faces correspond a null rotation.
Once all defect points are identified, desclination lines are reconstructed with a cluster analysis.

\section{Acknowledgements}

L.C.H.\ and D.A.B.\ were supported by the National
Science Foundation under Grant No.\ DMR-2225543. I. P. acknowledges the support from Ministerio de Ciencia, Innovaci\'on y Universidades MCIU/AEI/FEDER for financial support under Grant Agreement No. PID2024-156516NB-100 AEI/FEDER-EU, and Generalitat de Catalunya for financial support under Program Icrea Acad\`emia and Project No. 2021SGR-673. 
G.N. acknowledges EPSRC for access to the HPC resources at EPCC (Cirrus).

\bibliographystyle{unsrt}  

\bibliography{biblio}

@article{marchetti1993translational,
  author  = {Marchetti, M. Cristina and Nelson, David R.},
  title   = {Translational Correlations in the Vortex Array at the Surface of a Type-{II} Superconductor},
  journal = {Physical Review B},
  volume  = {47},
  pages   = {12214--12223},
  year    = {1993},
  doi     = {10.1103/PhysRevB.47.12214}
}

@article{balents1993delocalization,
  author  = {Balents, Leon and Kardar, Mehran},
  title   = {Delocalization of Flux Lines from Extended Defects by Bulk Randomness},
  journal = {Europhysics Letters},
  volume  = {23},
  number  = {7},
  pages   = {503--508},
  year    = {1993},
  doi     = {10.1209/0295-5075/23/7/007}
}

@article{teo2010topological,
  author  = {Teo, Jeffrey C. Y. and Kane, C. L.},
  title   = {Topological Defects and Gapless Modes in Insulators and Superconductors},
  journal = {Physical Review B},
  volume  = {82},
  pages   = {115120},
  year    = {2010},
  doi     = {10.1103/PhysRevB.82.115120}
}

@article{doostmohammadi2018,
  title={Active nematics},
  author={A. Doostmohammadi and M. Ign{\'e}s-Mullol and J.M. Yeomans and F. Sagu{\'e}s},
  journal={Nat. Commun.},
  volume={9},
  number={1},
  pages={3246},
  year={2018}
}

@Article{yashunksy2024,
author ="Yashunsky, V. and Pearce, D.J.G. and Ariel, G. and Be’er, A.",
title  ="Topological defects in multi-layered swarming bacteria",
journal  ="Soft Matter",
year  ="2024",
volume  ="20",
issue  ="21",
pages  ="4237-4245",
publisher  ="The Royal Society of Chemistry",
doi  ="10.1039/D4SM00038B",
url  ="http://dx.doi.org/10.1039/D4SM00038B"
}

@article{keber2014,
author = {F.C. Keber  and E. Loiseau  and T. Sanchez  and S.J. DeCamp  and L. Giomi  and M.J. Bowick  and M.C. Marchetti  and Z. Dogic  and A.R. Bausch },
title = {Topology and dynamics of active nematic vesicles},
journal = {Science},
volume = {345},
number = {6201},
pages = {1135-1139},
year = {2014},
doi = {10.1126/science.1254784},
URL = {https://www.science.org/doi/abs/10.1126/science.1254784},
eprint = {https://www.science.org/doi/pdf/10.1126/science.1254784}}

@article{armengol2023,
  title={Epithelia are multiscale active liquid crystals},
  author={J.M. Armengol-Collado and L.N. Carenza and J. Ecker and D. Krommydas and L. Giomi},
  journal={Nat. Phys.},
  volume={19},
  pages={1773--1779},
  year={2023}
}

@article{chiang2023,
author = {M. Chiang  and A. Hopkins  and B. Loewe  and M.C. Marchetti  and D. Marenduzzo },
title = {Intercellular friction and motility drive orientational order in cell monolayers},
journal = {PNAS},
volume = {121},
number = {40},
pages = {e2319310121},
year = {2024}
}

@article{cates1987,
  title={Reptation of living polymers: dynamics of entangled polymers in the presence of reversible chain-scission reactions},
  author={M.E. Cates},
  journal={Macromolecules},
  volume={20},
  number={9},
  pages={2289--2296},
  year={1987},
  publisher={ACS Publications}
}

@article{marchetti,
  author = {M.C. Marchetti and J.F. Joanny and S. Ramaswamy and  T.B. Liverpool and J. Prost and  M. Rao and R.A. Simha},
  title = {Hydrodynamics of soft active matter},
  journal = {Rev. Mod. Phys.},
  volume = {85},
  pages = {1143},
  year = {2013}
}

@ARTICLE{tjhung,
  author = {E. Tjhung and D. Marenduzzo and M.E. Cates},
  title = {Spontaneous symmetry breaking in active droplets provides a generic route to motility},
  journal = {PNAS},
  volume = {109},
  pages = {12381-12386},
  year = {2012},
}

@article{copar2013,
  title={Quaternions and hybrid nematic disclinations},
  author={S. {\v{C}}opar  and S. {\v{Z}}umer},
  journal={Proc. R. Soc. A},
  volume={469},
  number={2156},
  pages={20130204},
  year={2013},
  publisher={The Royal Society Publishing}
}

@article{copar2014,
  title={Topology and geometry of nematic braids},
  author={S. {\v{C}}opar},
  journal={Phys. Rep.},
  volume={538},
  number={1},
  pages={1--37},
  year={2014},
  publisher={Elsevier},
}

@Article{hemingway2016,
author ="Hemingway, E.J. and Mishra, P. and Marchetti, M.C. and Fielding, S.M.",
title  ="Correlation lengths in hydrodynamic models of active nematics",
journal  ="Soft Matter",
year  ="2016",
volume  ="12",
issue  ="38",
pages  ="7943-7952",
publisher  ="The Royal Society of Chemistry"}

@article{duclos2020,
  title={Topological structure and dynamics of three-dimensional active nematics},
  author={Duclos, G. and Adkins, R. and Banerjee, D. and Peterson, M.S.E. and Varghese, M. and Kolvin, I. and Baskaran, A. and Pelcovits, R.A. and Powers, T.R and Baskaran, A. and Toschi, F. and Hagan, M.F. and Streichan, S.J. and Vitelli, V. and Beller, D.A. and Dogic, Z.},
  journal={Science},
  volume={367},
  number={6482},
  pages={1120--1124},
  year={2020},
  publisher={American Association for the Advancement of Science}
}

@article{tjhung2015,
  title={A minimal physical model captures the shapes of crawling cells},
  author={Tjhung, E. and Tiribocchi, A. and Marenduzzo, D. and Cates, M.E.},
  journal={Nat. Commun.},
  volume={6},
  number={1},
  pages={5420},
  year={2015}
}

@article{sanchez2012,
  title={Spontaneous motion in hierarchically assembled active matter},
  author={Sanchez, T. and Chen, D.T.N. and DeCamp, S.J. and Heymann, M. and Dogic, Z.},
  journal={Nature},
  volume={491},
  number={7424},
  pages={431--434},
  year={2012}
}

@article{carenza2019,
  title={Rotation and propulsion in 3D active chiral droplets},
  author={Carenza, L.N. and Gonnella, G. and Marenduzzo, D. and Negro, G.},
  journal={PNAS},
  volume={116},
  number={44},
  pages={22065--22070},
  year={2019},
  publisher={National Acad Sciences}
}

@article{binysh2020,
  title={Three-dimensional active defect loops},
  author={Binysh, J. and Kos, {\v{Z}}. and {\v{C}}opar, S. and Ravnik, M. and Alexander, G.P.},
  journal={Phys. Rev. Lett.},
  volume={124},
  number={8},
  pages={088001},
  year={2020}
}

@article{ruske2021,
  title={Morphology of active deformable 3D droplets},
  author={Ruske, L. and Yeomans, J.M.},
  journal={Phys. Rev. X},
  volume={11},
  number={2},
  pages={021001},
  year={2021},
  publisher={APS}
}

@article{sulaiman,
  title={Lattice Boltzmann algorithm to simulate isotropic-nematic emulsions},
  author={Sulaiman, N. and Marenduzzo, D. and Yeomans, J. M.},
  journal={Phys. Rev. E},
  volume={74},
  pages={041708},
  year={2006},
  publisher={APS}
}

@article{hatwalne,
  title={Rheology of Active-Particle Suspensions},
  author={Hatwalne, Y. and Ramaswamy, S. and Rao, M. and Aditi Simha, R.},
  journal={Phys. Rev. Lett.},
  volume={92},
  pages={118101},
  year={2004},
  publisher={APS}
}

@article{kumar2014,
  title={Actomyosin contractility rotates the cell nucleus},
  author={Kumar, A. and Maitra, A. and Sumit, M. and Ramaswamy, S. and Shivashankar, G.V.},
  journal={Sci. Rep.},
  volume={4},
  number={1},
  pages={3781},
  year={2014}
}

@article{schimming2022,
  title={Singularity identification for the characterization of topology, geometry, and motion of nematic disclination lines},
  author={Schimming, C.D. and Vi{\~n}als, J.},
  journal={Soft Matter},
  volume={18},
  number={11},
  pages={2234--2244},
  year={2022},
  publisher={Royal Society of Chemistry}
}

@article{giomi,
  title={Defect dynamics in active nematics},
  author={Giomi, L. and Bowick, M.J. and Mishra, P. and Sknepnek, R. and Marchetti, M.C.},
  journal={Phil. Trans. Roy. Soc. A},
  year={2014},
  volume={372},
  pages={2029},
}

@book{degennes,
  title={The Physics of Liquid Crystals},
  author={de Gennes, P.G. and Prost, J.},
  year={1993},
  publisher={Clarendon Press, Oxford, 2nd edn}
}

@book{beris,
  title={Thermodynamics of Flowing Systems},
  author={Beris, A.N. Edwards, B.J.},
  year={1994},
  publisher={Oxford University Press, Oxford}
}

@article{ramachandran2011mayavi,
  title={{Mayavi: 3D Visualization of Scientific Data}},
  author={Ramachandran, P. and Varoquaux, G.},
  journal={Computing in Science \& Engineering},
  volume={13},
  number={2},
  pages={40--51},
  issn={1521-9615},
  year={2011},
  publisher={IEEE}
}

@article{kralj2023,
  title = {Defect Line Coarsening and Refinement in Active Nematics},
  author = {Kralj, N. and Ravnik, M. and Kos, \ifmmode \check{Z}\else \v{Z}.},
  journal = {Phys. Rev. Lett.},
  volume = {130},
  issue = {12},
  pages = {128101},
  numpages = {6},
  year = {2023},
  doi = {10.1103/PhysRevLett.130.128101},
  url = {https://link.aps.org/doi/10.1103/PhysRevLett.130.128101}
}

@article{shendruk2018,
  title = {Twist-induced crossover from two-dimensional to three-dimensional turbulence in active nematics},
  author = {Shendruk, T.N. and Thijssen, K. and Yeomans, J.M. and Doostmohammadi, A.},
  journal = {Phys. Rev. E},
  volume = {98},
  issue = {1},
  pages = {010601},
  numpages = {7},
  year = {2018}
}

@article{Nejad2023,
  title = {Spontaneous Rotation of Active Droplets in Two and Three Dimensions},
  author = {Nejad, M.R. and Yeomans, J.M.},
  journal = {PRX Life},
  volume = {1},
  issue = {2},
  pages = {023008},
  numpages = {13},
  year = {2023}
}

@article {Armengol2024,
article_type = {journal},
title = {Hydrodynamics and multiscale orderin confluent epithelia},
author = {J.M. Armengol-Collado and Carenza, L.N. and Giomi, L.},
editor = {Goldstein, Raymond E},
volume = 13,
year = 2024,
pub_date = {2024-01-08},
pages = {e86400},
citation = {eLife 2024;13:e86400},
doi = {10.7554/eLife.86400},
url = {https://doi.org/10.7554/eLife.86400},
journal = {eLife},
issn = {2050-084X}
}

@article{Hoffmann2022,
   author = {L.A. Hoffmann and  L.N. Carenza and J. Eckert and L. Giomi},
   issue = {15},
   journal = {Sci. Adv.},
   pages = {eabk2712},
   title = {Theory of defect-mediated morphogenesis},
   volume = {8},
   url = {https://www.science.org/doi/abs/10.1126/sciadv.abk2712},
   year = {2022},
}

@article{martinez2019,
  title={Selection mechanism at the onset of active turbulence},
  author={Mart{\'\i}nez-Prat, B. and Ign{\'e}s-Mullol, J. and Casademunt, J. and Sagu{\'e}s, F.},
  journal={Nat. Phys.},
  volume={15},
  number={4},
  pages={362--366},
  year={2019}
}

@article{hardouin2019,
  title={Reconfigurable flows and defect landscape of confined active nematics},
  author={Hardo{\"u}in, J. and Hughes, R. and Doostmohammadi, A. and Laurent, J. and Lopez-Leon, T. and Yeomans, J.M. and Ign{\'e}s-Mullol, J. and Sagu{\'e}s, F.},
  journal={Commun. Phys.},
  volume={2},
  number={1},
  pages={121},
  year={2019}
}

@Article{shendruk2017,
author ="Shendruk, T.N. and Doostmohammadi, A. and Thijssen, K. and Yeomans, J.M.",
title  ="Dancing disclinations in confined active nematics",
journal  ="Soft Matter",
year  ="2017",
volume  ="13",
issue  ="21",
pages  ="3853-3862",
doi  ="10.1039/C6SM02310J"
}

@article{Giomi2015,
   author = {L. Giomi},
   issue = {3},
   journal = {Phys. Rev. X},
   pages = {31003},
   title = {Geometry and Topology of Turbulence in Active Nematics},
   volume = {5},
   year = {2015},
}

@article{simha2002,
  title = {Hydrodynamic Fluctuations and Instabilities in Ordered Suspensions of Self-Propelled Particles},
  author = {Simha, R.A. and Ramaswamy, S.},
  journal = {Phys. Rev. Lett.},
  volume = {89},
  issue = {5},
  pages = {058101},
  numpages = {4},
  year = {2002}
}

@article{Alert2022,
   author = {R. Alert and J. Casademunt and J.-F. Joanny},
   doi = {10.1146/annurev-conmatphys-082321-035957},
   issue = {1},
   journal = {Annu. Rev. Condens. Matter Phys.},
   pages = {143-170},
   title = {Active Turbulence},
   volume = {13},
   url = {https://doi.org/10.1146/annurev-conmatphys-082321-035957},
   year = {2022},
}

@article{tran2016,
author = {L. Tran  and M.O. Lavrentovich  and D.A. Beller  and N. Li  and K.J. Stebe  and R.D. Kamien },
title = {Lassoing saddle splay and the geometrical control of topological defects},
journal = {PNAS},
volume = {113},
number = {26},
pages = {7106-7111},
year = {2016},
doi = {10.1073/pnas.1602703113},
URL = {https://www.pnas.org/doi/abs/10.1073/pnas.1602703113},
eprint = {https://www.pnas.org/doi/pdf/10.1073/pnas.1602703113}
}

@BOOK{mermin1990,
   author = {{Mermin}, N.D.},
    title = "{Boojums All the Way Through}",
     year = 1990,
    place={Cambridge}, 
    publisher={Cambridge University Press}, 
   adsurl = {https://ui.adsabs.harvard.edu/abs/1990bawt.book.....M}
}

@article{turner2010,
  title = {Vortices on curved surfaces},
  author = {Turner, A.M. and Vitelli, V. and Nelson, D.R.},
  journal = {Rev. Mod. Phys.},
  volume = {82},
  issue = {2},
  pages = {1301--1348},
  numpages = {0},
  year = {2010}
}

@article{Liu2013,
  title={Nematic liquid crystal boojums with handles on colloidal handlebodies},
  author={Liu, Q. and Bohdan, S. and Tasinkevych, M. and Smalyukh, I.I.},
  journal={PNAS},
  volume={110},
  number={23},
  pages={9231-9236},
  year={2013}
}

@article{ellis2018,
  title={Curvature-induced defect unbinding and dynamics in active nematic toroids},
  author={Ellis, P.W. and Pearce, D.J.G. and Chang, Y.-W. and Goldsztein, G. and Giomi, L. and Fernandez-Nieves, A.},
  journal={Nat. Phys.},
  volume={14},
  number={1},
  pages={85--90},
  year={2018}
}

@article{friedel1969,
author = {Friedel, J. and De Gennes, P.G.},
title = {Buckling due to distortion in liquid crystals},
journal = {CR Acad. Sc. Paris B},
volume = {268},
pages = {257-259},
year = {1969},
}

@article{marenduzzo2007,
  title = {Steady-state hydrodynamic instabilities of active liquid crystals: Hybrid lattice Boltzmann simulations},
  author = {Marenduzzo, D. and Orlandini, E. and Cates, M.E. and Yeomans, J.M.},
  journal = {Phys. Rev. E},
  volume = {76},
  issue = {3},
  pages = {031921},
  numpages = {18},
  year = {2007},
  doi = {10.1103/PhysRevE.76.031921},
  url = {https://link.aps.org/doi/10.1103/PhysRevE.76.031921}
}

@article{Doostmohammadi2017,
   author = {A. Doostmohammadi and T.N. Shendruk and K. Thijssen and J.M. Yeomans},
   doi = {10.1038/ncomms15326},
   journal = {Nat. Commun.},
   title = {Onset of meso-scale turbulence in active nematics},
   volume = {8},
   url = {http://gen.lib.rus.ec/scimag/index.php?s=10.1038/ncomms15326},
   year = {2017},
}

@article{head2024-3,
  title={Majorana quasiparticles and topological phases in 3D active nematics},
  author={Head, L.C. and Negro, G. and Carenza, L.N. and Johnson, N. and Keogh, R.R. and Gonnella, G. and Morozov, A. and Orlandini, E. and Shendruk, T.N. and Tiribocchi, A. and Marenduzzo, D.},
  journal={PNAS},
  volume={121},
  number={52},
  pages={e2405304121},
  year={2024},
  publisher={National Academy of Sciences}
}

@article{opathalage2019,
author = {A. Opathalage  and M.M. Norton  and M.P.N. Juniper  and B. Langeslay  and S.A. Aghvami  and S. Fraden  and Z. Dogic },
title = {Self-organized dynamics and the transition to turbulence of confined active nematics},
journal = {PNAS},
volume = {116},
number = {11},
pages = {4788-4797},
year = {2019},
doi = {10.1073/pnas.1816733116},
URL = {https://www.pnas.org/doi/abs/10.1073/pnas.1816733116},
eprint = {https://www.pnas.org/doi/pdf/10.1073/pnas.1816733116}}

@Article{l_head,
author={Head, L.C.
and Dor{\'e}, C.
and Keogh, R.R.
and Bonn, L.
and Negro, G.
and Marenduzzo, D.
and Doostmohammadi, A.
and Thijssen, K.
and L{\'o}pez-Le{\'o}n, T.
and Shendruk, T.N.},
title={Spontaneous self-constraint in active nematic flows},
journal={Nat. Phys.},
volume={20},
pages={492--500},
year={2024},
issn={1745-2481},
doi={10.1038/s41567-023-02336-5},
url={https://doi.org/10.1038/s41567-023-02336-5}
}

@article{copar2019,
  title = {Topology of Three-Dimensional Active Nematic Turbulence Confined to Droplets},
  author = {\ifmmode \check{C}\else \v{C}\fi{}opar, S. and Aplinc, J. and Kos, \ifmmode \check{Z}\else \v{Z}. and \ifmmode \check{Z}\else \v{Z}\fi{}umer, S. and Ravnik, M.},
  journal = {Phys. Rev. X},
  volume = {9},
  issue = {3},
  pages = {031051},
  numpages = {13},
  year = {2019}
}

@article{napoli2020,
  title = {Spontaneous helical flows in active nematics lying on a cylindrical surface},
  author = {Napoli, G. and Turzi, S.},
  journal = {Phys. Rev. E},
  volume = {101},
  issue = {2},
  pages = {022701},
  numpages = {6},
  year = {2020}
}

@article{maroudas2021,
  title={Topological defects in the nematic order of actin fibres as organization centres of Hydra morphogenesis},
  author={Maroudas-Sacks, Y. and Garion, L. and Shani-Zerbib, L. and Livshits, A. and Braun, E. and Keren, K.},
  journal={Nat. Phys.},
  volume={17},
  number={2},
  pages={251--259},
  year={2021},
  publisher={Nature Publishing Group UK London}
}

@book{stewart2004,
  title={The static and dynamic continuum theory of liquid crystals: a mathematical introduction.},
  author={Stewart, I.W.},
  year={2004},
  publisher={Taylor \& Francis},
  address={London}
}

@book{deGennes1979,
  title={Scaling Concepts in Polymer Physics},
  author={de Gennes, P.G.},
  year={1979},
  publisher={Cornell University Press},
  address={Ithaca, NY}
}

@article{blow2014,
  title = {Biphasic, Lyotropic, Active Nematics},
  author = {Blow, M.L. and Thampi, S.P. and Yeomans, J.M.},
  journal = {Phys. Rev. Lett.},
  volume = {113},
  issue = {24},
  pages = {248303},
  numpages = {5},
  year = {2014},
  publisher = {American Physical Society},
  doi = {10.1103/PhysRevLett.113.248303},
  url = {https://link.aps.org/doi/10.1103/PhysRevLett.113.248303}
}

@article{vafa2024,
  title = {Periodic orbits, pair nucleation, and unbinding of active nematic defects on cones},
  author = {Vafa, F. and Nelson, D.R. and Doostmohammadi, A.},
  journal = {Phys. Rev. E},
  volume = {109},
  issue = {6},
  pages = {064606},
  numpages = {15},
  year = {2024},
  publisher = {American Physical Society},
  doi = {10.1103/PhysRevE.109.064606},
  url = {https://link.aps.org/doi/10.1103/PhysRevE.109.064606}
}

@article{zhang2016,
	author = {Zhang, R. and Zhou, Y. and Rahimi, M. and de Pablo, J.J.},
	date = {2016/11/21},
	doi = {10.1038/ncomms13483},
	id = {Zhang2016},
	isbn = {2041-1723},
	journal = {Nat. Commun.},
	number = {1},
	pages = {13483},
	title = {Dynamic structure of active nematic shells},
	url = {https://doi.org/10.1038/ncomms13483},
	volume = {7},
	year = {2016}}

@article{Guillamat2016,
  title = {\textcolor{black}{Probing the shear viscosity of an active nematic film}},
  author = {Guillamat, P. and Ign\'es-Mullol, J. and Shankar, S. and Marchetti, M.C. and Sagu\'es, F.},
  journal = {Phys. Rev. E},
  volume = {94},
  issue = {6},
  pages = {060602},
  numpages = {5},
  year = {2016}
  }

@article{Digregorio2024,
  title = {Coexistence of Defect Morphologies in Three-Dimensional Active Nematics},
  author = {Digregorio, P. and Rorai, C. and Pagonabarraga, I. and Toschi, F.},
  journal = {Phys. Rev. Lett.},
  volume = {132},
  issue = {25},
  pages = {258301},
  numpages = {6},
  year = {2024}
}

@article{negro2025,
	author = {Negro, G. and Head, L.C. and Carenza, L.N. and Shendruk, T.N. and Marenduzzo, D. and Gonnella, G. and Tiribocchi, A.},
	doi = {10.1038/s41467-025-56236-8},
	journal = {Nat. Commun.},
	number = {1},
	pages = {1412},
	title = {Topology controls flow patterns in active double emulsions},
	volume = {16},
	year = {2025},}

@article{Chandragiri2020,
  title = {Flow States and Transitions of an Active Nematic in a Three-Dimensional Channel},
  author = {Chandragiri, S. and Doostmohammadi, A. and Yeomans, J.M. and Thampi, S.P.},
  journal = {Phys. Rev. Lett.},
  volume = {125},
  issue = {14},
  pages = {148002},
  numpages = {6},
  year = {2020}
}

@article{Alaimo2017,
  title = {Curvature controlled defect dynamics in topological active nematics},
  author = {Alaimo, F. and K\"{o}hler, C. and Voigt, A.},
  journal = {Sci. Rep.},
  volume = {7},
  pages = {5211},
  year = {2017}
}

@article{keogh2022,
  title = {Helical flow states in active nematics},
  author = {Keogh, R.R. and Chandragiri, S. and Loewe, B. and Ala-Nissila, T. and Thampi, S.P. and Shendruk, T.N.},
  journal = {Phys. Rev. E},
  volume = {106},
  issue = {1},
  pages = {L012602},
  numpages = {6},
  year = {2022}
}

@article{micheletti2011,
title = {Polymers with spatial or topological constraints: Theoretical and computational results},
journal = {Phys. Rep.},
volume = {504},
number = {1},
pages = {1-73},
year = {2011},
issn = {0370-1573},
doi = {https://doi.org/10.1016/j.physrep.2011.03.003},
url = {https://www.sciencedirect.com/science/article/pii/S0370157311000640},
author = {C. Micheletti and D. Marenduzzo and E. Orlandini}
}

@misc{argento_gliomas_2025,
    title = {Gliomas organize as liquid crystals: three-dimensional nematic order, disclinations and quasi-long-range order},
    shorttitle = {Gliomas organize as liquid crystals},
    publisher = {bioRxiv},
    author = {Argento, A.E. and Varela, M.L. and Singh, G. and Visnuk, D.P. and Jacobovitz, B. and Rutherford, M.E. and Edwards, M.B. and Chaboche, Q. and Orringer, D.A. and Heth, J.A. and Castro, M.G. and Beller, D.A. and Blanch-Mercader, C. and Lowenstein, P.R.},
    year = {2025}
}

@book{bernevig2013,
  title={Topological insulators and topological superconductors},
  author={Bernevig, B.A.},
  year={2013},
  publisher={Princeton University Press}
}

@article{shankar_defect_2018,
    title = {Defect {Unbinding} in {Active} {Nematics}},
    volume = {121},
    doi = {10.1103/PhysRevLett.121.108002},
    number = {10},
    urldate = {2024-06-21},
    journal = {Phys. Rev. Lett.},
    author = {Shankar, S. and Ramaswamy, S. and Marchetti, M.C. and Bowick, M.J.},
    year = {2018},
    pages = {108002},
}

@article{kralj_chirality_2024,
    title = {Chirality, anisotropic viscosity and elastic anisotropy in three-dimensional active nematic turbulence},
    volume = {7},
    copyright = {2024 The Author(s)},
    issn = {2399-3650},
    url = {https://www.nature.com/articles/s42005-024-01720-8},
    doi = {10.1038/s42005-024-01720-8},
    language = {en},
    number = {1},
    urldate = {2024-07-08},
    journal = {Commun. Phys.},
    author = {Kralj, N. and Ravnik, M. and Kos, Ž.},
    year = {2024},
    note = {Publisher: Nature Publishing Group},
    pages = {1--9},
}

\end{document}